%% file: ms.tex
\documentclass[fleqn,usenatbib]{mnras}

\usepackage{natbib}

\usepackage[T1]{fontenc}

\DeclareRobustCommand{\VAN}[3]{#2}
\let\VANthebibliography\thebibliography
\def\thebibliography{\DeclareRobustCommand{\VAN}[3]{##3}\VANthebibliography}

\usepackage{graphicx}	
\usepackage{amsmath}	
\usepackage{amssymb}	
\usepackage[para]{threeparttablex}

\newcommand{\ha}{H$\upalpha$}
\newcommand{\hb}{H$\upbeta$}

\newcommand{\hi}{H\,\textsc{i}}

\newcommand{\ppxf}{\textsc{ppxf}}

\newcommand{\forb}[2]{[#1\,\textsc{#2}]}
\newcommand{\eso}{ESO\,422--G028}

\graphicspath{{figs/}}

\title[A neutral inflow and star formation in ESO\,422--G028]{Revisiting the Giant Radio Galaxy ESO\,422--G028: Part I. Discovery of a neutral inflow and recent star formation in a restarted giant}

\author[H. R. M. Zovaro et al.]{Henry R. M. Zovaro$^{1}$\thanks{E-mail: henry.zovaro@anu.edu.au},
	Chris J. Riseley$^{2,3,4}$,
	Philip Taylor$^{1,6}$,
	Nicole P. H. Nesvadba$^{7}$,
	Tim J. Galvin$^{4,5}$,
	\newauthor
	Umang Malik$^{1}$, Lisa J. Kewley$^{1,6}$\\
$^{1}$Research School of Astronomy and Astrophysics, The Australian National University, Canberra, ACT 2611, Australia\\
$^{2}$Dipartimento di Fisica e Astronomia, Universit\`{a} degli Studi di Bologna, via P. Gobetti 93/2, 40129 Bologna, Italy\\
$^{3}$INAF-Istituto di Radioastronomia, via P. Gobetti, 101, 40129 Bologna, Italy\\
$^{4}$CSIRO Astronomy and Space Science, PO Box 1130, Bentley, WA 6102, Australia\\
$^{5}$International Centre for Radio Astronomy Research, Curtin University, Bentley, WA 6102, Australia\\
$^{6}$ARC Centre of Excellence for All Sky Astrophysics in 3 Dimensions (ASTRO 3D)\\
$^{7}$Universit\'e de la C\^{o}te d'Azur, Observatoire de la C\^{o}te d'Azur, CNRS, Laboratoire Lagrange, Bd de l'Observatoire, CS 34229, 06304 Nice cedex 4, France\\
}

\date{Accepted XXX. Received YYY; in original form ZZZ}

\pubyear{2021}

\usepackage{newtxtext,newtxmath}
\begin{document}
\label{firstpage}
\pagerange{\pageref{firstpage}--\pageref{lastpage}}
\maketitle

\begin{abstract}
\input{abstract.tex}
\end{abstract}




\input{body.tex}

\section*{Acknowledgements}

The authors thank T. Mendel, G. V. Bicknell, A. Y. Wagner and D. Mukherjee for helpful discussions that improved this work.
CJR acknowledges financial support from the ERC Starting Grant `DRANOEL', number 714245 and UM is supported by the Australian Government Research Training Program (RTP) Scholarship. 
LJK and HRMZ gratefully acknowledge the support of an ARC Laureate Fellowship (FL150100113).

The bulk of this research was carried out at Mount Stromlo Observatory, on the traditional lands of the Ngunnawal and Ngambri people, and the observations presented in this work were gathered at Siding Spring Observatory, which is located on the traditional lands of the Gamilaraay/Kamilaroi people.

This research has made use of the NASA/IPAC Extragalactic Database, which is funded by the National Aeronautics and Space Administration and operated by the California Institute of Technology.

This research made use of the \textsc{python} packages \textsc{Matplotlib}\footnote{\url{https://matplotlib.org/}}\,\citep{Hunter2007},  \textsc{NumPy}\footnote{\url{https://numpy.org/}}\,\citep{Harris2020}, \textsc{SciPy}\footnote{\url{http://www.scipy.org/}}\,\citep{Scipy2001}, and \textsc{Astropy},\footnote{\url{http://www.astropy.org}} a community-developed core package for Astronomy\,\citep{Astropy2013,Astropy2018}. 

This work used the DiRAC@Durham facility managed by the Institute for Computational Cosmology on behalf of the STFC DiRAC HPC Facility (\url{www.dirac.ac.uk}). The equipment was funded by BEIS capital funding via STFC capital grants ST/P002293/1, ST/R002371/1 and ST/S002502/1, Durham University and STFC operations grant ST/R000832/1. DiRAC is part of the National e-Infrastructure.

Stellar models generated with \textsc{Sed@.0} code\footnote{\textsc{Sed@} is a synthesis code included in the {\it Legacy Tool project} of the {\it Violent Star Formation Network}; see {\it \textsc{Sed@} Reference Manual} at \url{http://www.iaa.es/~mcs/sed@} for more information.} with the following inputs: IMF from \cite{Salpeter1955} in the mass range $0.1-120 \rm\, M_\odot$; High Resolution library from \citet{Martins2004,GonzalezDelgado2005} based on atmosphere models from \textsc{Phoenix} \citep{Hauschildt&Baron1999,Allard2001}, \textsc{Atlas9} \citep{Kurucz1991} computed with \textsc{Spectrum} \citep{Gray&Corbally1994}, \textsc{Atlas9} library computed with \textsc{Synspec} \citep{Hubeny&Lanz2011}, and \textsc{Tlusty} \citep{Lanz&Hubeny2003}. 
Geneva isochrones computed with the isochrone program presented in \cite{Meynet1995} and following the prescriptions quoted in \cite{Cervino2001} from the evolutionary tracks from \cite{Schaller1992} at $ Z=0.001$/$ Z=0.020$; \cite{Charbonnel1993} at $ Z=0.004$; \cite{Schaerer1993a} at $ Z=0.008$ and \cite{Schaerer1993b} at $ Z=0.040$.
Padova isochrones presented in \cite{Girardi2002}\footnote{Available at \url{http://pleiadi.pd.astro.it/}} based on the (solar scaled mixture) tracks from \cite{Girardi2000,Bertelli1994} that includes overshooting and a simple synthetic evolution of TP-AGB \cite{Girardi&Bertelli1998}.

\section*{Data Availability}

The data underlying this article will be shared on reasonable request to the corresponding author.



\bibliographystyle{mnras}
\bibliography{bibliography.bib}



\appendix
\input{appendix.tex}

%


\bsp	
\label{lastpage}
\end{document}

%% file: abstract.tex
Giant radio galaxies provide important clues into the life cycles and triggering mechanisms of radio jets.
With large-scale jets spanning 1.8 Mpc, \eso{} ($z = 0.038$) is a giant radio galaxy that also exhibits signs of restarted jet activity in the form of pc-scale jets.
We present a study of the spatially-resolved stellar and gas properties of \eso{} using optical integral field spectroscopy from the WiFeS spectrograph.
In addition to the majority $\sim 13\,\rm Gyr$ old stellar population, \eso{} exhibits a much younger ($\lesssim 10\,\rm Myr$ old) component with an estimated mass of $ 10^{7.6}\,\rm M_\odot$ which is predominantly located in the North-West region of the galaxy.
Unusually, the ionised gas kinematics reveal two distinct disks traced by narrow ($\sigma_{\rm H\upalpha} < 100 \,\rm km\,s^{-1}$) and broad ($\sigma_{\rm H\upalpha} > 150 \,\rm km\,s^{-1}$) \ha{} emission respectively. Both ionised gas disks are misaligned with the axis of stellar rotation, suggesting an external origin.
This is consistent with the prominent interstellar Na\,D absorption, which traces a $1 - 3 \,\rm M_\odot \, yr^{-1}$ inflow of neutral gas from the North.
We posit that an inflow of gas -- either from an accretion event or a gas-rich merger -- has triggered both the starburst and the restarted jet activity, and that \eso{} is potentially on the brink of an epoch of powerful AGN activity.

%% file: body.tex
\defcitealias{Subrahmanyan2008}{S08}

\section{Introduction}

\begin{figure}
	\centering
	\includegraphics[width=\linewidth]{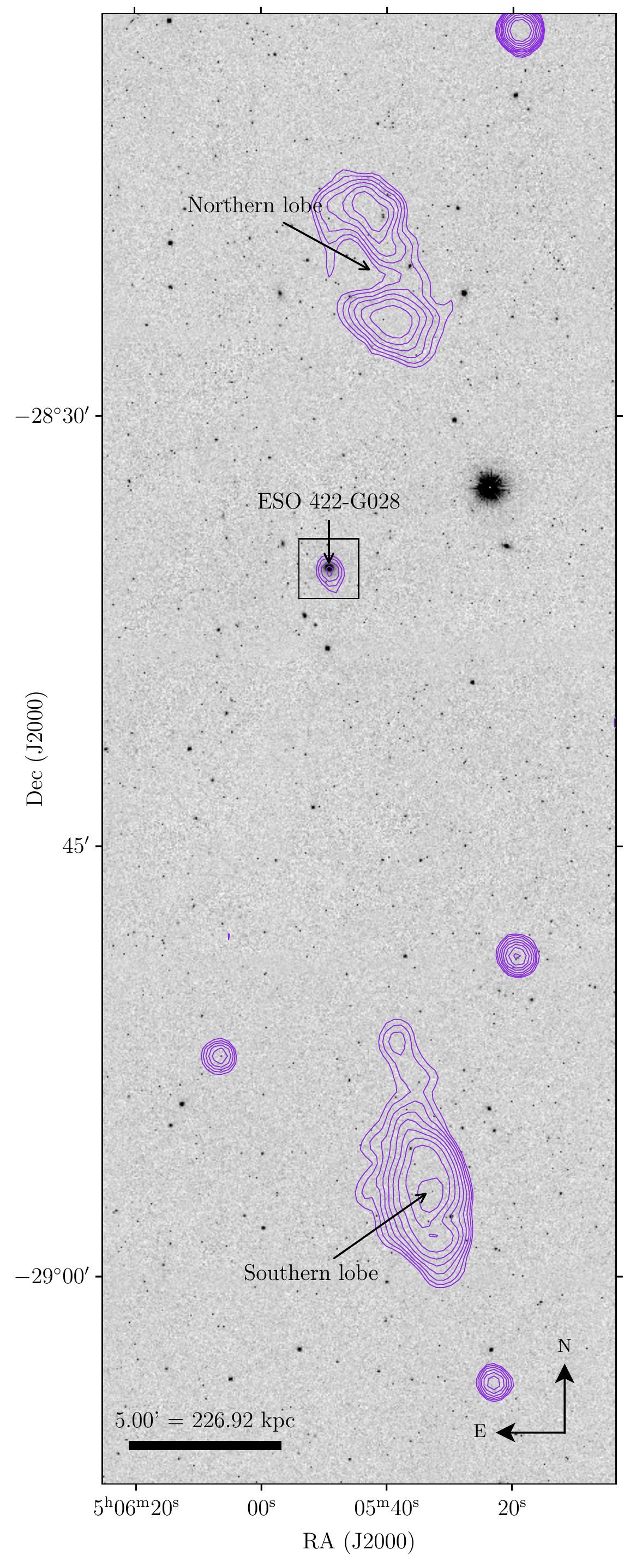}
	\caption{\textit{VISTA} Hemisphere Survey $K$-band image~\citep{McMahon2013} showing \eso{} and its large-scale environment. The colour scaling of the image has been adjusted to enhance visibility of \eso{}. The contours show the 1.4 GHz NRAO VLA Sky Survey (NVSS) image~\citep{Condon1998}, revealing the Mpc-scale radio lobes, and represent 10 logarithmically spaced intervals from $3\,\rm mJy\,beam^{-1}$ to $50\,\rm mJy\,beam^{-1}$. The box surrounding \eso{} shows the FoV of Fig.~\ref{fig: FoV (small)}.}
	\label{fig: FoV (large)}
\end{figure}

The largest single structures in the known Universe, Giant Radio Galaxies (GRGs) are radio galaxies with jets spanning distances greater than 0.7\,Mpc in extent.
First discovered by \citet{Willis1974}, these giants are uncommon, with fewer than 1000 catalogued~\citep{Delhaize2021} and only a handful studied in detail. 
Typically hosted by early-type galaxies (ETGs), the largest known to-date is approximately 5 Mpc in size~\citep{Machalski2008}, although only $15\,\rm per \, cent$ are larger than 2 Mpc~\citep{Dabhade2020a}.
A combination of long-lived or high-power periods of AGN activity~\citep{Subrahmanyan1996,Wiita1989} and low-density local environments~\citep{Mack1998,Malarecki2015} have been proposed as potential mechanisms enabling the jets to reach their gargantuan sizes.

In recent years, low-frequency radio surveys such as the The MeerKAT International GHz Tiered Extragalactic Exploration~\citep[MIGHTEE;][]{Jarvis2016} and the LOFAR Two-Metre Sky Survey~\citep[LoTSS;][]{Shimwell2017} have led to the discovery of hundreds of GRGs~\citep{Dabhade2020a,Dabhade2020b,Delhaize2021}, facilitating population studies of these rare objects. 
Despite the influx of new data, the precise conditions required to form these giant radio jets remains unclear.
For example, \citet{Lan&Prochaska2021} showed that GRGs do \textit{not} preferentially reside in lower-density environments in comparison to ``regular'' radio galaxies; indeed, \citet{Dabhade2020a} found GRGs in field, group and cluster environments. 
Meanwhile, \citet{Dabhade2020b} found that GRGs tend to have lower Eddington ratios than their regular-sized counterparts, but similar supermassive black hole masses, suggesting the radiative efficiency of the accretion disk may be key to producing powerful jets.
Additionally, roughly 5 per cent of GRGs are ``double-double'' sources with two sets of jets from different epochs of AGN activity~\citep{Dabhade2020a}, suggesting that shortly-spaced episodes of AGN fuelling may also be important.
Detailed studies of individual GRGs -- including the properties of the host galaxy -- are therefore crucial in elucidating the mechanisms that contribute to the growth of the jets.
In particular, studying the fuel sources of GRGs may help us to figure out what is so special about these giants, and give us clues as to their life cycle.

Absorption lines are particularly useful for detecting inflows powering the jet activity, because gas can only be observed in absorption \textit{in front of} the stellar continuum. As a result, inflows and outflows can be unambiguously identified, in contrast to emission lines: redshifted absorption traces gas flowing into the galaxy, and blueshifted absorption traces gas flowing out of the galaxy. 
At optical wavelengths, interstellar absorption by neutral sodium -- the Na\,D doublet at $5889.9 \, \text{\AA}$ and $5889.9\,\text{\AA}$ -- can be used to detect inflows and outflows of neutral gas.

Although an estimated $1/3$ of radio galaxies at $z < 0.2$ exhibit interstellar Na\,D absorption~\citep{Lehnert2011}, in most cases it traces outflowing material. Similarly, \hi{} is also most often observed in outflows~\citep{Morganti2005,Morganti&Oosterloo2018}.
However, inflows of neutral gas traced by \hi{} have been reported in radio galaxies~\citep{vanGorkom1989} and may fuel AGN activity. Moreover, a recent study by \citet{Roy2021b} found that redshifted interstellar Na\,D absorption appears to be common in quiescent intermediate-mass radio galaxies, suggesting such inflows can trigger the AGN.

The present work is part of a series of papers on the nearby GRG \eso{}~\citep[$z = 0.038150$;][]{Jones2009}, which exhibits lobes spanning approximately 1.8 Mpc that are no longer powered by the central engine~\citep[][hereafter S08]{Subrahmanyan2008}. \eso{} also exhibits evidence for restarted jet activity in the form of pc-scale jets (Riseley et al., \textit{in prep.}).
In this paper we use high spectral resolution integral field spectroscopy to study the spatially resolved stellar and gas properties of \eso{}.
Our observations reveal a young stellar population in this GRG, plus complex ionised gas kinematics as traced by \ha{} emission. We also detect prominent redshifted Na\,D absorption that traces an inflow, potentially fuelling both star formation and the restarted jet activity.

For the remainder of this paper we assume a flat $\Lambda \rm CDM$ cosmology with $\Omega_{\rm M} = 0.3$, $\Omega_\Lambda = 0.7$ and $H_0 = 70\,\rm km\,s^{-1}\,Mpc^{-1}$. The coordinates and distances of \eso{} assumed in this work are given in Table~\ref{tab: ESO422 properties}.

\section{The Giant Radio Galaxy \eso{}}

Independently discovered to be a GRG in 1986 by both \citet{Saripalli1986} and \citet{Subrahmanya&Hunstead1986}, \eso{} is a nearby giant elliptical galaxy with  radio lobes spanning a projected linear extent of 1.8\,Mpc, with properties summarised in Table~\ref{tab: ESO422 properties}.
As shown in Fig.~\ref{fig: FoV (large)}, the radio source comprises two prominent FRII-type lobes extending roughly North-South and unresolved core emission centred on the host.
The lobes are diffuse, with no prominent hotspots, and have ill-defined boundaries, indicating they are no longer being powered by the jet, with an estimated spectral age of 0.3\,Gyr~\citepalias{Subrahmanyan2008}.

Unusually for a GRG, the lobes are quite asymmetric, with a length ratio of 1:1.6. The Northern jet also bends to the West. At low frequencies, the backflow from the Northern lobe extends all the way back to the host, whilst that from the Southern lobe only reaches three quarters of the distance back towards the host~\citep[see fig. 8 of ][]{Beardsley2019}.
The Northern lobe exhibits a dramatic steepening of the spectral index towards its Northern boundary, contrary to what is typically observed in edge-brightened radio lobes, which \citetalias{Subrahmanyan2008} speculate is due to interaction with the more dense ambient medium in the North.

\begin{figure}
	\centering
	\includegraphics[width=\linewidth]{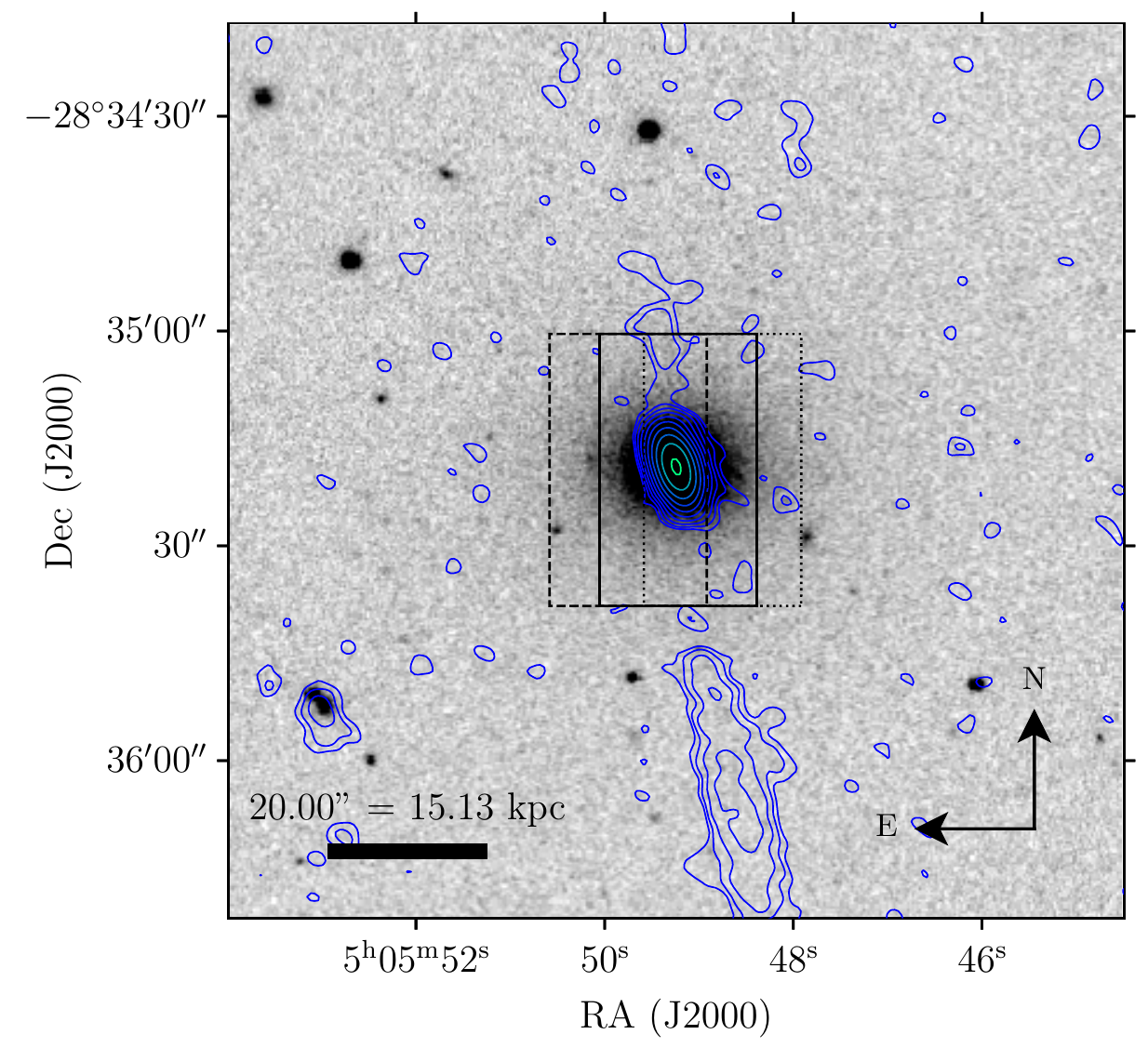}
	\caption[]{A zoom-in of the \textit{VISTA} Hemisphere Survey $K$-band image~\citep{McMahon2013} indicated by the black box in Fig.~\ref{fig: FoV (large)}. The solid, dashed and dotted boxes indicate the fields of view of our three different WiFeS pointings, and the contours show the 4.85~GHz VLA image representing 10 logarithmically-spaced intervals from $0.075\,\rm mJy\,beam^{-1}$ to $5.5\,\rm mJy\,beam^{-1}$.
	}
	\label{fig: FoV (small)}
\end{figure}

The kpc-resolution Very Large Array (VLA) image of \citetalias{Subrahmanyan2008}, overlaid in Fig.~\ref{fig: FoV (small)}, shows the collimated kpc-scale Southern jet and unresolved core emission. The relative brightness of the Southern jet compared to the Northern lobe indicates the Southern lobe is inclined towards the Earth.
The VLA observations (project AS262, taken October 19 and 20 1986 in CnB configuration; see \citetalias{Subrahmanyan2008} for details) were sourced from the NRAO VLA Archive Survey\footnote{ \url{http://archive.nrao.edu/nvas/}} and re-imaged using \textsc{WSclean}~\citep{Offringa2014} version 2.10.0\footnote{ \url{https://sourceforge.net/p/wsclean/wiki/Home/}}. We employed \texttt{Briggs} weighting with \texttt{robust}~$=+1.0$~\citep{Briggs1995} and used multi-scale clean, with scales between 0 (point sources) and $12\times$ the theoretical beam size, in order to more accurately reconstruct extended radio emission from the jets of \eso{}. Our final VLA image has a representative off-source rms noise of $0.025\rm \, mJy \, beam^{-1}$, where the restoring beam is $7.58\times3.96$~arcsec at $\rm{PA} = 18.25^\circ$, and a reference frequency of 4.85~GHz.

\begin{table}
	\centering
	\caption{Properties of ESO\,422--G028 assumed in this paper.}
	\begin{threeparttable}
		\begin{tabular}{cc}
			\hline
			\textbf{Property} & \textbf{Value} \\
			\hline
			RA (J2000) & $\rm 05^h05^m49\overset{s}{.}22$\tnote{a}\\
			Dec (J2000) & $\rm-28^\circ35^\prime19.4^{\prime\prime}$\tnote{a}\\
			Redshift ($z$) & $0.038150$\tnote{b}\\
			Heliocentric recessional velocity & $11437 \,{\rm km\,s^{-1}}$\tnote{b}\\
			Angular diameter distance ($D_{\rm A}$) & $156.02 \,\rm Mpc$\\
			Luminosity distance ($D_{\rm L}$) & $168.15 \,\rm Mpc$\\
			\hline
			Stellar mass ($M_*$) & $10^{11.4} \,{\rm M_\odot}$\tnote{c} \\
			BH mass ($M_{\rm BH}$) & $1.7^{+0.4}_{-0.3} \times 10^9 \,{\rm M_\odot}$\tnote{c} \\
			Dust mass ($M_{\rm dust}$) & $\sim 10^6 - 10^7 \,{\rm M_\odot}$\tnote{d} \\
			Effective radius ($R_e$) & $12.56 \, {\rm arcsec}$\tnote{e} \\
			Star formation rate (SFR) & $0.39 \pm 0.01 {\rm \, M_\odot \, yr^{-1}}$\tnote{c} \\
			Radio luminosity at 1.4\,GHz ($L_{1.4\rm\,GHz}$) & $1.07 \times 10^{25} \,{\rm W \, Hz^{-1}}$\tnote{f} \\		
			\hline
		\end{tabular}
		\begin{tablenotes}
			\footnotesize
			\item[a]\citet{Jarrett2000}.\item[b]\citet{Jones2009}.\item[c]This work.\item[d]\citet{Trifalenkov1994}.\item[e]\citet{Lauberts&Valentijn1989}.\item[f]\citetalias{Subrahmanyan2008}.
		\end{tablenotes}
		\label{tab: ESO422 properties}
	\end{threeparttable}
\end{table}

\subsection{Environment}

\eso{} is the most massive galaxy in a small group of at least 5 members. The group has a projected diameter of 0.3\,Mpc and a low velocity dispersion, indicating it is unvirialised~\citepalias[$\sigma = 83\,\rm km\,s^{-1}$;][]{Subrahmanyan2008}.
The small group is located approximately 1.8\,Mpc from a large, dense filament of galaxies spanning North-East, which has previously been identified as a supercluster by \citet{Kalinkov&Kuneva1995}; the region to the South-West is sparse in comparison.
\citetalias{Subrahmanyan2008} estimate the group is falling towards the supercluster at $\sim 600 \,\rm km\,s^{-1}$.

The pronounced bend in the Northern jet, as well as the strong asymmetry in the two jet lengths, has been attributed to buoyant forces arising from the motion of the group towards the supercluster, as well as higher ambient densities near the large-scale structure~\citepalias{Subrahmanyan2008}.
This latter explanation is supported by the excess of soft X-ray sources in the North~\citep{Jamrozy2005}. Similar phenomena have been observed in other GRGs~\citep{Saripalli1986}.
\citetalias{Subrahmanyan2008} also posited that galactic superwinds from AGN in the supercluster may contribute to the asymmetry between the two jets.

\section{Observations and data reduction}

\eso{} was observed with the Wide-Field Spectrograph \citep[WiFeS; ][]{Dopita2007,Dopita2010} on the Australian National University 2.3\,m telescope at Siding Spring Observatory.
WiFeS is an image slicer integral field spectrograph comprising 25 1''-width slices with 0.5'' sampling along the image, binned to 1'' to better match the seeing, providing 1''\,$\times$\,1''  spatial pixels (\textit{spaxels}) over a 25''\,$\times$ 38'' field-of-view. 

\eso{} was observed on the 25th and 26th of November 2019 (PI Riseley, proposal ID 4190027) and on the 20th of November 2020 (PI Malik, proposal ID 4200140) using the low-resolution B3000 (3200--5000\,\AA, $R \sim 3000$, $\Delta v \approx 100 \,\rm km\,s^{-1}$) and the high-resolution R7000 (5290--7060\,\AA, $R \sim 7000$, $\Delta v \approx 40 \,\rm km\,s^{-1}$) gratings with the RT560 beam splitter. Three pointings were used to cover the full extent of the galaxy, as shown in Fig.~\ref{fig: FoV (small)}. Observations were dithered in $\pm1$'' intervals North-South and East-West to mitigate the effects of bad pixels, each with a 1200\,s exposure time for a total effective exposure time of 15600\,s. 
The star HD36702 was used as flux and telluric standard~\citep{Bessell1999}.

The observations were reduced in the standard way using \textsc{Pywifes}, the  data reduction pipeline for WiFeS~\citep{Childress2014}.
Cu-Ar and Ne-Ar arc lamp exposures were used to derive the wavelength solution, and exposures of the coronagraphic wire mask were used to calibrate the spatial alignment of the slits.
Quartz lamp and twilight flat exposures were used to correct for wavelength and spatial variations in the instrument response respectively.
Standard star exposures were used to correct for telluric absorption and for flux calibration.
Two data cubes were generated for each exposure, corresponding to the blue (B3000) and red (R7000) arms of the spectrograph respectively.

Sky subtraction was carried out following the method of \citet{Zovaro2020}. Regions within the field-of-view with no source signal were used to estimate the sky spectrum, which was then subtracted from each data cube. Sky subtraction residuals were minimised by scaling the flux of the sky spectrum to the measured intensity of a subset of sky lines in each spaxel. 
The instrumental resolution was estimated from the widths of sky lines to be $\rm FWHM_{\rm B3000} = 1.4$\,\AA{} and $\rm FWHM_{\rm R7000} = 0.9$\,\AA{} in the B3000 and R7000 data cubes respectively.

Mosaics were created by spatially shifting the data cubes by eye and combining them by taking the sigma-clipped mean of each pixel. $1\sigma$ errors for each pixel value in the final mosaic were derived by combining the variance extensions from each individual cube.
The mosaicked data cubes were corrected for foreground Galactic extinction assuming $A_V = 0.0454$ from the extinction map of \citet{Schlafly&Finkbeiner2011} and using the reddening curve of \citet{Fitzpatrick&Massa2007} with $R_V = 3.1$.

\begin{figure*}
	\includegraphics[width=1\linewidth]{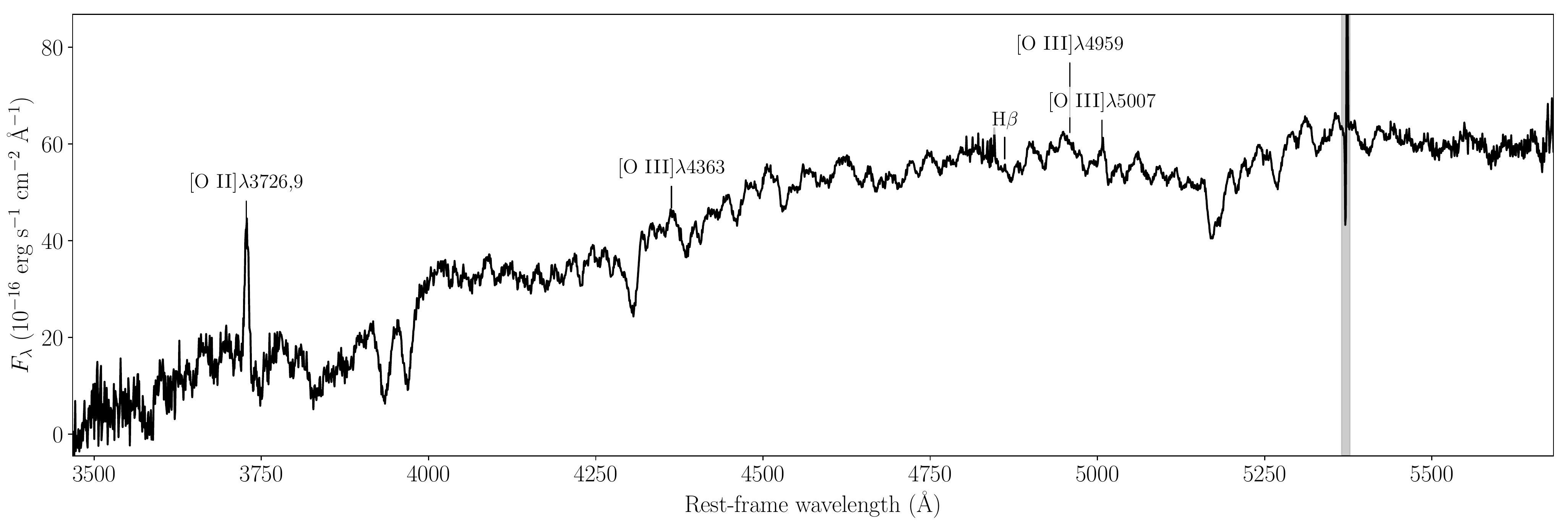}
	\includegraphics[width=1\linewidth]{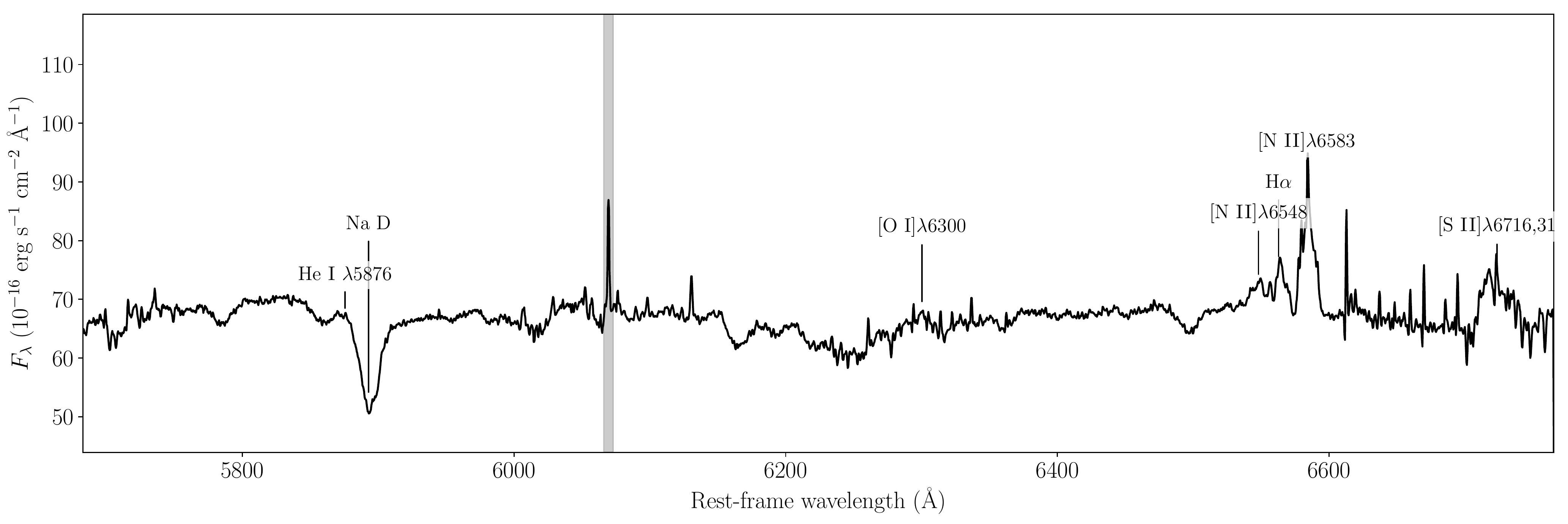}
	\caption{Integrated spectra of \eso{} from the blue (top) and red (bottom) WiFeS data cubes, extracted from an aperture with radius $1R_e$. The shaded grey regions denote spectral windows dominated by sky line residuals. The R7000 spectrum has been smoothed with a boxcar filter with radius of 3 spectral pixels.}
	\label{fig: integrated spectra}
\end{figure*}

Fig.~\ref{fig: integrated spectra} shows the spectra from the B3000 and R7000 data cubes extracted from an aperture of radius $1 R_e$ centred on \eso{}. 
The continuum is dominated by deep stellar absorption features, superimposed with strong \ha{}, \forb{N}{ii} and \forb{S}{ii} emission exhibiting complex, asymmetric line profiles. The comparative paucity of the \hb{} and \forb{O}{iii} lines firmly classifies \eso{} as a low-excitation radio galaxy~\citep[LERG;][]{BestHeckman2012}.

The pseudo $V$-band continuum, generated by summing the wavelength slices from 5000--6000\,\AA, is shown in Fig.~\ref{fig: Voronoi bins, V-band continuum & line profiles}. The surface brightness profile is smooth; no morphological perturbations are apparent at the modest 1'' spatial resolution of WiFeS. 
Also shown are \ha{} and \forb{N}{ii} line profiles extracted from the Eastern, central and Western regions of the galaxy, and a position-velocity (PV) diagram which shows the line profiles extracted from a horizontal slit across the centre of the galaxy. Although regular rotation is apparent in the PV diagram, the line profiles are highly asymmetric and vary dramatically across the galaxy.

\begin{figure}
	\centering
	\includegraphics[width=0.8\linewidth]{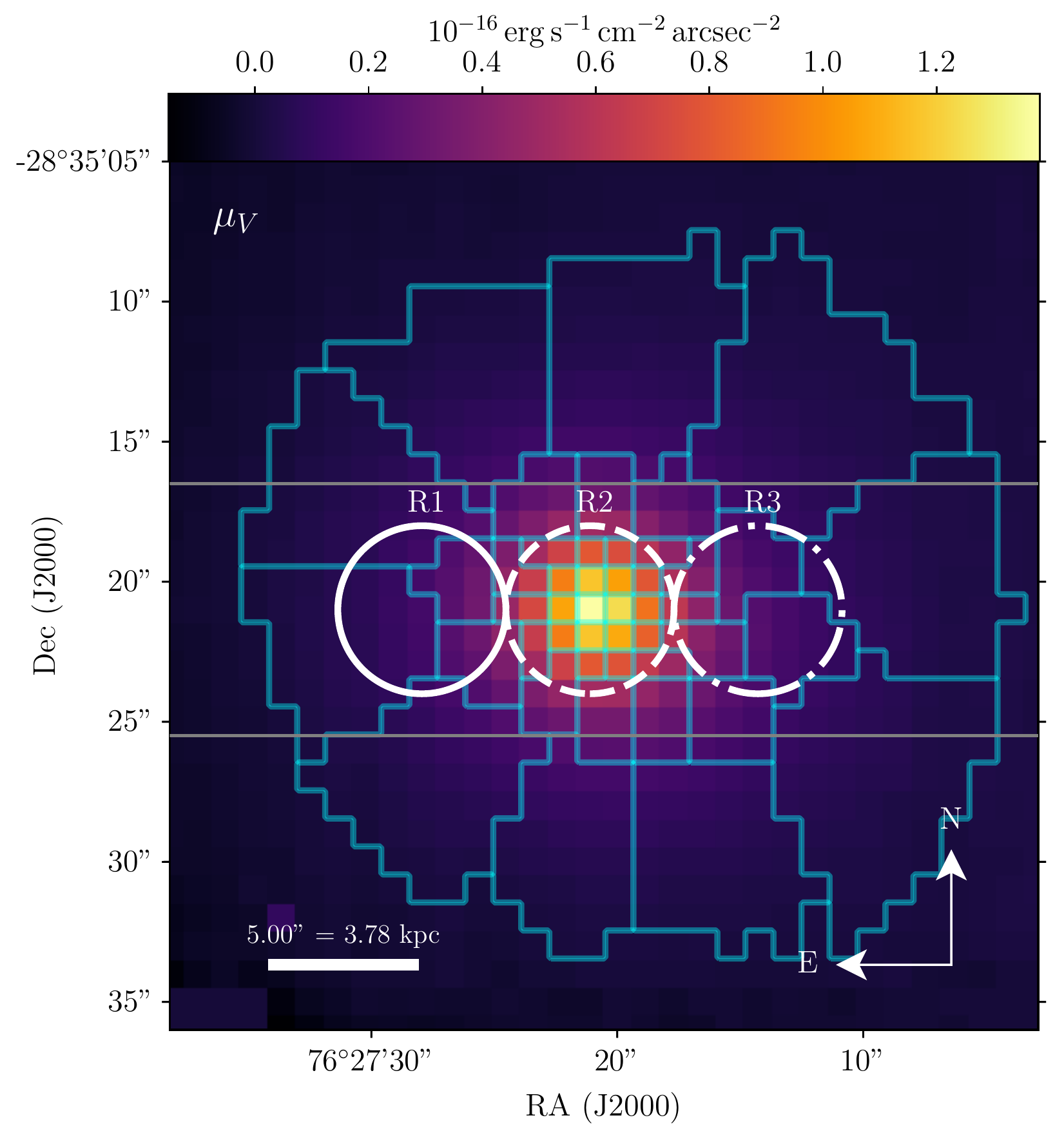}
	\includegraphics[width=1\linewidth]{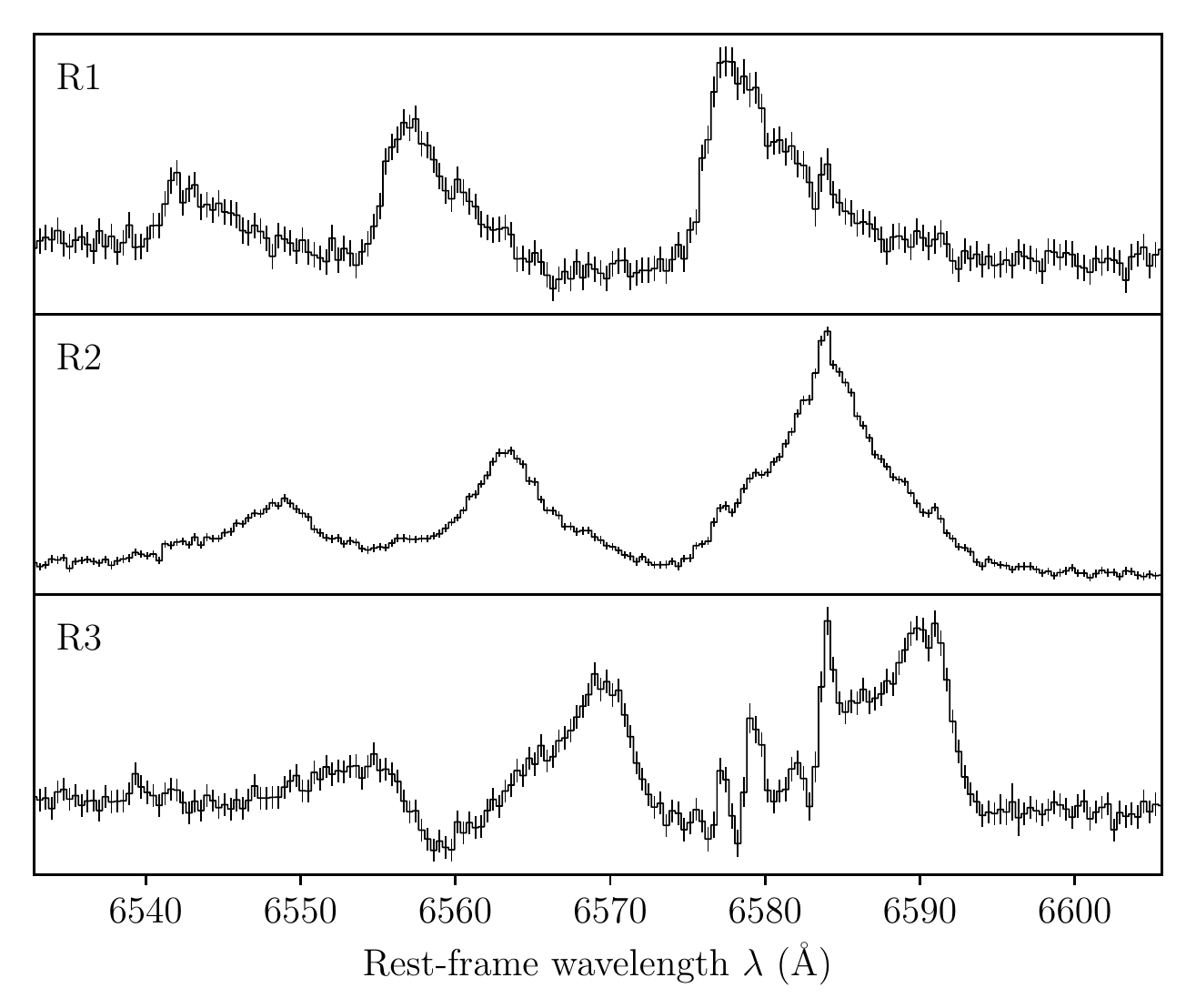}
	\includegraphics[width=1\linewidth]{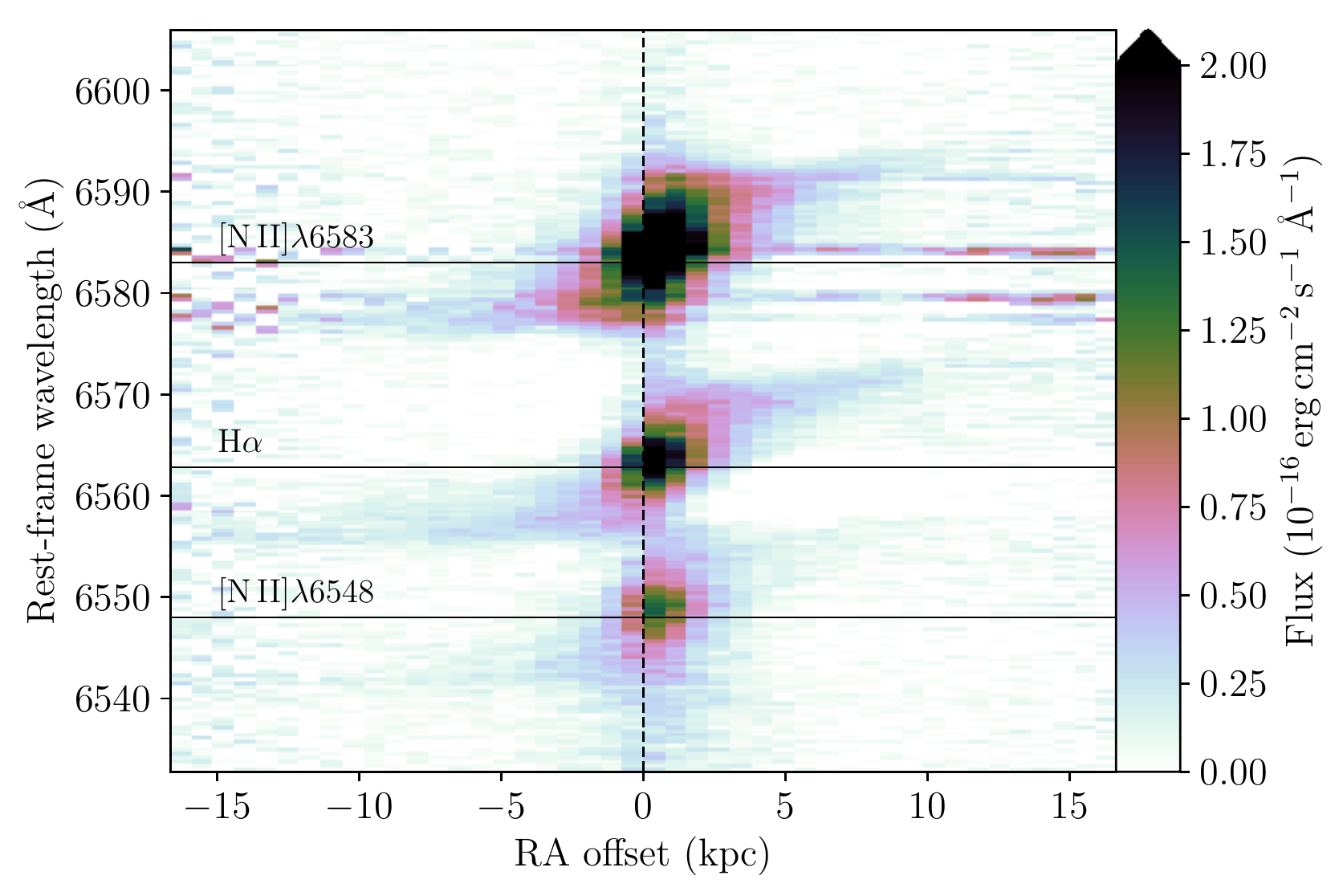}
	\caption{
		The top panel shows the pseudo $V$-band continuum generated by summing the WiFeS data cube from 5000--6000\,\AA , overlaid with the boundaries of the Voronoi bins in cyan.
		Spectra extracted from the apertures R1, R2 and R3, centred on the \ha{} and \forb{N}{ii} line profiles, are shown in the middle panel, with fluxes scaled arbitrarily. The narrow features visible in the R3 spectrum are skyline residuals.
		The bottom panel shows a PV diagram running East-West, also centred on the \ha{} and \forb{N}{ii} lines, extracted from the spaxels between the horizontal lines shown in the top panel. Rotation is clearly visible, as are the broad, asymmetric line profiles at larger radii.}
	\label{fig: Voronoi bins, V-band continuum & line profiles}
\end{figure}

Due to the steep surface brightness profile of \eso{}, only a handful of spaxels in the centre of the galaxy have sufficient S/N to reliably determine the properties of the stellar population and emission-line gas.
We therefore spatially binned the data cube using a Voronoi tessellation via the \textsc{python} implementation of \textsc{VorBin}\footnote{\url{https://www-astro.physics.ox.ac.uk/~mxc/software/\#binning}}~\citep{Cappellari&Copin2003}. This binning method enforces a minimum S/N ratio in each bin whilst ensuring round, compact bin shapes. To determine the binning, we generated an image and the corresponding variance by summing the B3000 data and variance cubes from 5280--5288\,\AA, and enforced a minimum S/N of 60. This translated to a median S/N between 10 and 25 per spectral pixel for each bin. Fig.~\ref{fig: Voronoi bins, V-band continuum & line profiles} shows the resulting bins. Binned data and variance cubes were generated by simply adding the spectra from each spaxel of the data and variance cubes respectively; these binned cubes were used for the remainder of the analysis.

\section{Stellar population analysis}\label{sec: Stellar population analysis}

We used the \textsc{python} implementation of Penalized Pixel Fitting~\citep[\ppxf{};][]{Cappellari&Emsellem2004,Cappellari2017}\footnote{ \url{https://www-astro.physics.ox.ac.uk/~mxc/software/\#ppxf}} to analyse the age, metallicity and kinematics of the stellar population in each bin by determining the linear combination of simple stellar population (SSP) templates that best describes the stellar continuum.

In order to fit the stellar continuum across the full wavelength range of our WiFeS observations, we spectrally convolved the R7000 cube with a Gaussian kernel with $\rm FWHM = \sqrt{FWHM_{\rm B3000}^2 - FWHM_{\rm R7000}^2}$ before spectrally binning the cube to the resolution of the B3000 cube and merging the cubes together, creating a ``combined'' data cube.

Visual inspection of the spectra revealed deep Na\,D absorption~\citep[rest-frame wavelengths $5889.9 \,\text{\AA}$ and $5895.9\,\text{\AA}$;][]{Morton1991}. In some bins, the two lines comprising the doublet are clearly resolved, and are much narrower than that expected from stellar absorption, indicating the absorption is partially interstellar in origin. We therefore carried out our \ppxf{} analysis with this feature masked out to prevent the interstellar absorption biasing the fit.
The interstellar Na\,D absorption is discussed at length in Section~\ref{sec: Na D absorption line analysis}.

We used the SSP templates of \citet{GonzalezDelgado2005} generated from the Padova isochrones with a Salpeter IMF, spanning an age range of approximately 4\,Myr to 18\,Gyr divided into 75 bins with three metallicities ($0.2\rm\, Z_\odot$, $0.4\,\rm Z_\odot$ and $0.95\,\rm Z_\odot$).
These templates were chosen due to their high spectral resolution, meaning we did not need to degrade the spectral resolution of our observations prior to running \ppxf{}.
The Padova isochrones were used as they include stellar evolution along the red giant branch, which is important for accurately modelling older stellar populations characteristic of ETGs such as \eso{}, unlike the Geneva isochrones.
\ppxf{} was run twice in each bin: a first time in order to determine the best-fit age and metallicity, and a second time to determine the stellar kinematics.

To obtain the best-fit age and metallicity template weights in each bin, the stellar continuum and the emission lines were fitted to the ``combined'' data cube simultaneously, adopting independent velocity components for each. One kinematic component was adopted for the stars, and two components were used for the emission lines, each modelled by Gaussian profiles. 
For all kinematic components, only the line-of-sight (LOS) velocity and velocity dispersion were fit. 
A 4th-order multiplicative polynomial was included in the fit to compensate for extinction and calibration errors.
\textit{Regularisation} was used to bias the template weights towards the smoothest solution consistent with the data using the method detailed in \citet{Boardman2017}

We note that the emission lines were only fit during the \ppxf{} fit to ensure accurate fitting of the stellar continuum. We did not use the \ppxf{} fit to \textit{analyse} the gas emission lines: we instead fit these separately after subtraction of the stellar continuum, as detailed in Section~\ref{sec: emission line analysis; subsec: emission line fitting}, in order to achieve greater control over the fitting process required for our analysis.

\begin{figure*}
	\includegraphics[width=\linewidth]{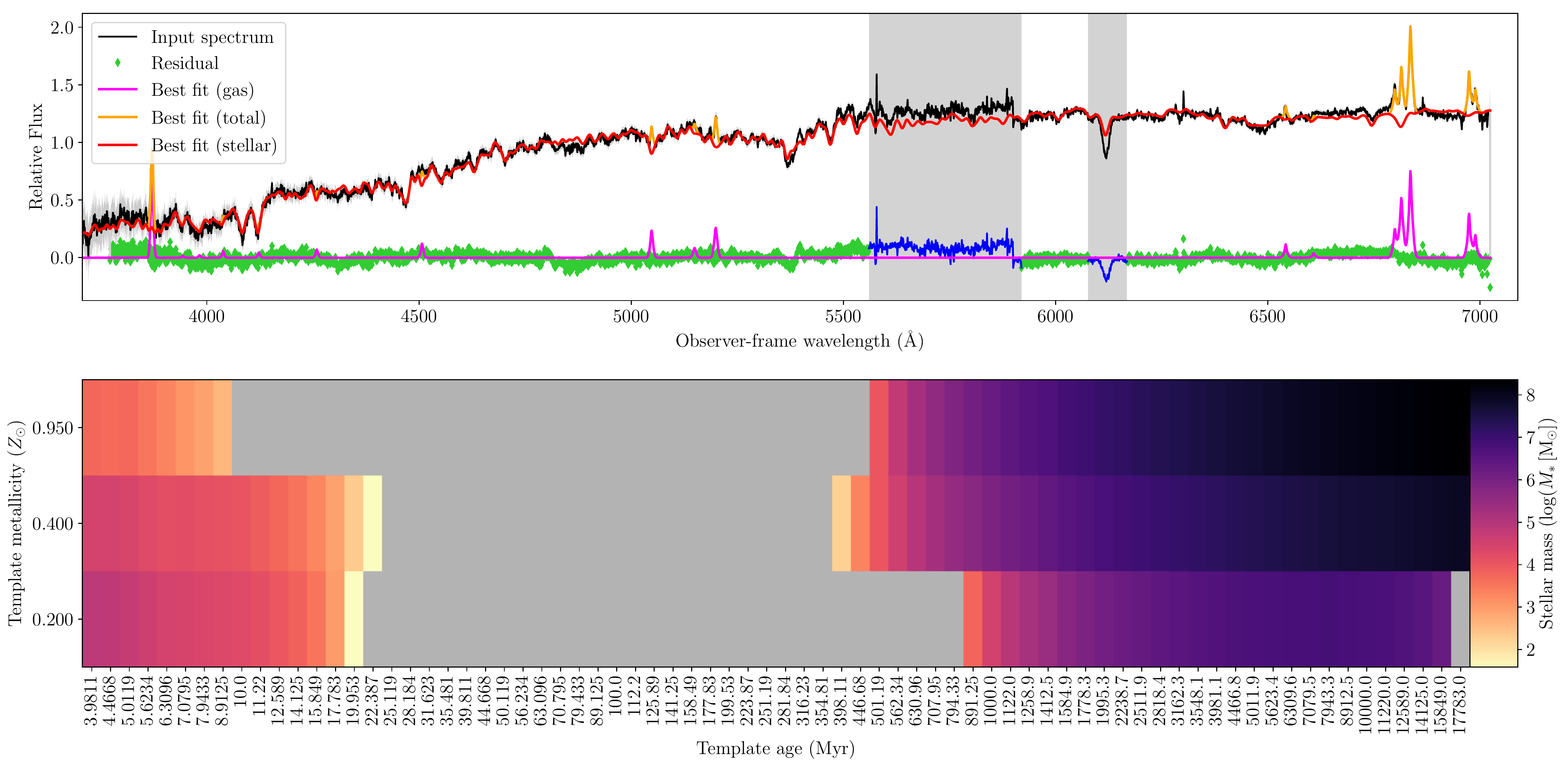}
	\caption{An example stellar continuum fit from one of the Voronoi bins. The upper plot shows the spectrum overlaid with the best-fit stellar continuum computed using \textsc{ppxf}; shaded regions in the spectrum denote regions masked out during the fit. The lower plot shows the corresponding mass-weighted stellar template weights, where the grey cells denote templates with a weight of zero.}
	\label{fig: example ppxf fits}
\end{figure*}

In all regions, \eso{} is dominated by old stellar populations; however, bins in the region to the North-West also exhibit a much younger component, with ages $\lesssim 10\,\rm Myr$. An example of the stellar fit to one such bin is shown in Fig.~\ref{fig: example ppxf fits}.

\begin{figure*}
	\includegraphics[width=1\linewidth]{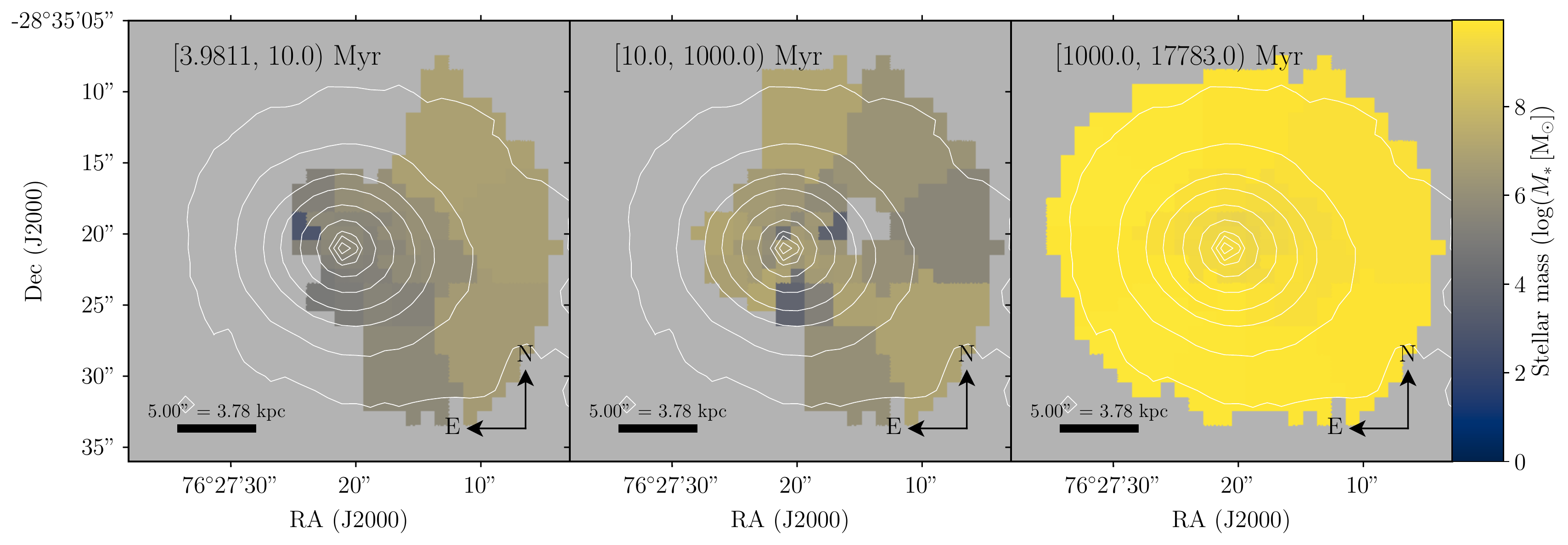}
	\includegraphics[width=0.95\linewidth]{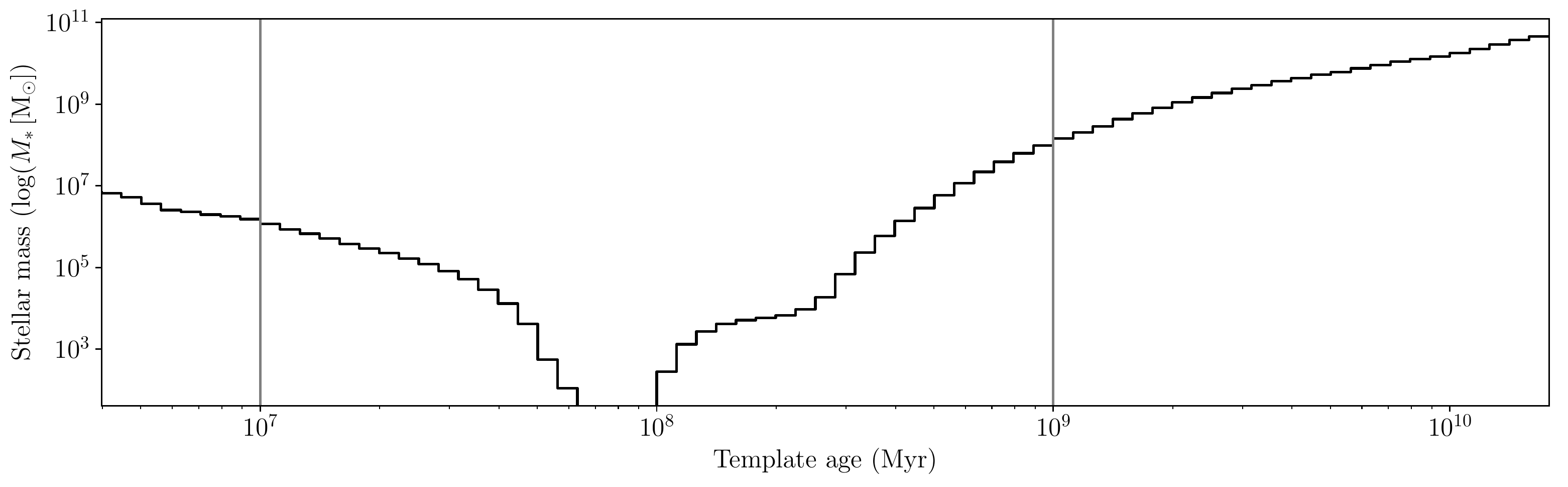}
	\caption{The stellar mass distribution from the \ppxf{} fit in each bin (top), divided into the age ranges indicated in the upper left corner. The contours represent the $V$-band continuum, with levels $2.5 \times 10^{-18}, \, 1 \times 10^{-17}, \, 2 \times 10^{-17}, \, 3 \times 10^{-17}, \, 5 \times 10^{-17}, \, 8 \times 10^{-17}, \, 1.1 \times 10^{-16}, \, 1.2 \times 10^{-16}$ and $1.3 \times 10^{-16}\,\rm erg \, s^{-1} \, cm^{-2} \, arcsec^{-2}$. The bottom panel shows the SFH of \eso{} integrated over all bins and template metallicities. The grey vertical lines show the cuts used to make the stellar mass maps in the upper panel.}
	\label{fig: stellar age map}
\end{figure*}

To estimate the total mass of both the old and young stellar populations, we converted the luminosity-weighted template weights into mass-weighted template weights. 
``Age slices'', shown in Fig.~\ref{fig: stellar age map}, show the total masses associated with stellar templates from the \ppxf{} fit within different age ranges in each bin. The young stellar populations are concentrated in the region to the North-West.
The weights and metallicities from all bins were added together to produce the histogram shown in Fig.~\ref{fig: stellar age map}, which essentially shows the star formation history (SFH) of \eso{}. There are two peaks in the SFH: one at $\lesssim \,\rm 10 \, Myr$, and another at $\gtrsim 10 \,\rm Gyr$.
The total mass of the young stellar population is approximately $10^{7.6}\,\rm M_\odot$, representing only a minute fraction of the total stellar mass, which is approximately $10^{11.4}\,\rm M_\odot$.

To confirm the presence of young stellar populations, we re-ran our \ppxf{} fits twice using (1) the Padova isochrones but with 1 kinematic component for the emission lines, and (2) the Geneva isochrones but with 2 kinematic components for the emission lines.
In every bin, the best-fit stellar continua and SFHs were very similar regardless of the number of kinematic components or the isochrones used, although our fits using the Geneva isochrones yielded a more metal-rich young stellar population a few Myr older than those predicted from our fits using the Padova isochrones. 
We therefore do not make any claims as to the precise metallicity or age of the young stellar population, except for that it has an age $\lesssim 10\,\rm Myr$.

\begin{figure*}
	\includegraphics[width=0.65\linewidth]{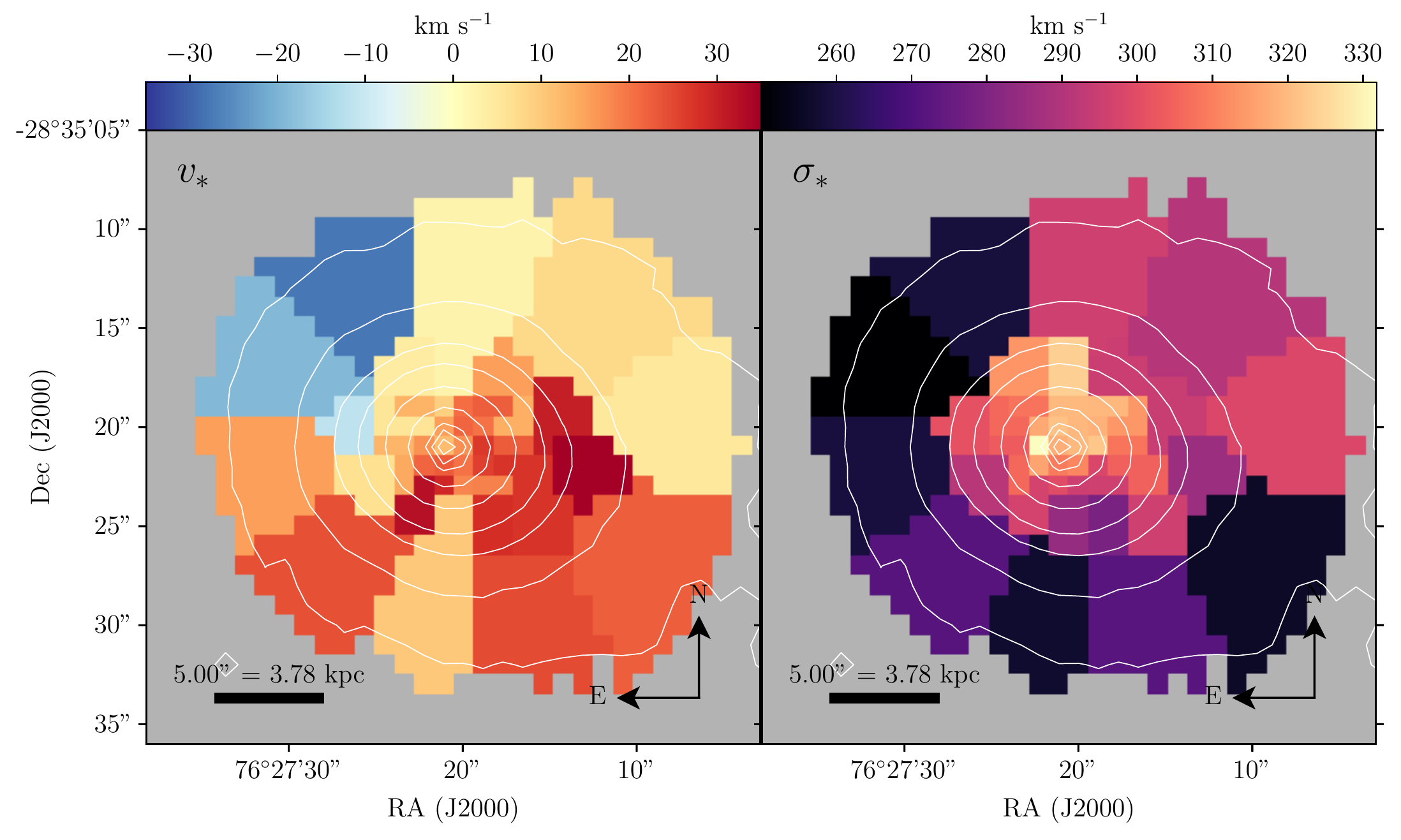}
	\caption{Stellar velocity, relative to systemic (left) and velocity dispersion (right) from the \ppxf{} fit. The contours are as in Fig.~\ref{fig: stellar age map}.}
	\label{fig: stellar kinematics}
\end{figure*}

The stellar kinematics were obtained using a separate \ppxf{} fit. First, the best-fit emission line profiles obtained from the age and metallicity fit were subtracted from the ``combined'' data cube. We then ran \ppxf{} on this cube with an additive 12th-order polynomial. Regularisation was not used in this fit, because it is not required to accurately constrain stellar kinematics. 
Fig.~\ref{fig: stellar kinematics} shows the LOS velocity and velocity dispersion. There is only weak stellar rotation, with velocities peaking at approximately $35\,\rm km\,s^{-1}$ relative to systemic, whereas the velocity dispersion is broad, reaching values of $330 \,\rm km\,s^{-1}$ in the central regions and decreasing smoothly outwards.
Using eqn.\,6 of \citet{Emsellem2007} we estimate a spin factor of $\lambda \sim 0.1$, classifying \eso{} as a slow rotator.

\section{Emission line analysis}\label{sec: emission line analysis; subsec: emission line fitting}

To analyse the emission lines in the binned spectra, an ``emission line-only'' data cube was first created by subtracting the best-fit stellar continuum in each bin from the binned data cube. 

Emission line fluxes were measured by simultaneously fitting the \forb{O}{ii}$\uplambda\uplambda3726,9\text{\AA\AA}$, \hb{}$\uplambda4861\text{\AA}$, \forb{O}{iii}$\uplambda\uplambda4959,5007\text{\AA\AA}$, \forb{O}{i}$\uplambda6300\text{\AA}$, \ha{}$\uplambda6563\text{\AA}$, \forb{N}{ii}$\uplambda\uplambda6548,83\text{\AA\AA}$ and \forb{S}{ii}$\uplambda\uplambda6716,31\text{\AA\AA}$ lines using \textsc{mpfit} \citep{Markwardt2009}, a \textsc{python} implementation of the Levenberg-Marquardt algorithm \citep{More1978} developed by M. Rivers\footnote{\url{http://cars9.uchicago.edu/software/python/mpfit.html}.}.
Multiple Gaussian components were fitted, where the number of required components was chosen by eye to account for the total flux of each line. Whilst most bins required one broad and one narrow component to achieve a satisfactory fit, others -- generally those with lower S/N -- only required a single component.
Within each kinematic component, every line was tied to the same radial velocity and velocity dispersion\footnote{Radial velocities were computed from the observed redshift $z_{\rm obs} = \uplambda_{\rm obs} / \uplambda_{\rm rest} - 1$ using $1 + z_{\rm obs} = (1 + z_{\rm cos})(1 + v_{\rm rad}/c)$.}, and the amplitude ratios of the lines in the \forb{N}{ii} and \forb{O}{iii} doublets were fixed to 3.06 and 2.94 respectively as dictated by quantum mechanics~\citep{Dopita&Sutherland2003}.
Rather than fit to the full spectrum, we only fitted to a narrow spectral window around each line in the range $\pm 600 \,\rm km\,s^{-1}$. In addition to the Gaussian components, a first-order polynomial was fitted within each of these windows to account for residuals due to errors in the stellar continuum fit.
Only those bins in which the best fit had a reduced-$\chi^2 < 2$ and $\rm S/N >3$ in the total fluxes of both \ha{} and \hb{} lines were kept.

We note that due to low S/N in the \hb{}, \forb{O}{iii} and \forb{O}{i} lines, the fluxes from the individual kinematic components are unreliable; however, care was taken during the fit to ensure the combined broad and narrow components accurately described the total emission line flux. 
This fitting process was only used to estimate total fluxes; we fit to the higher-S/N \ha{} and \forb{N}{ii} lines separately to more accurately determine the gas kinematics (see Section~\ref{subsec: emission line kinematics}).

Table~\ref{tab: emission line fluxes} shows the total emission line fluxes across all bins, corrected for extinction (see Section~\ref{subsec: extinction}).

\subsection{Extinction}\label{subsec: extinction}
$A_V$ was computed in each bin from the total \ha{} and \hb{} emission line fluxes using the reddening curve of \citet{Fitzpatrick&Massa2007} with $R_V = 3.1$ and assuming an intrinsic \ha{}/\hb{} flux ratio of 2.86~\citep{Dopita&Sutherland2003}. $A_V$ was only computed in those bins in which the S/N of the \ha{}/\hb{} ratio exceeded 3.

\begin{figure}
	\includegraphics[height=0.7\linewidth]{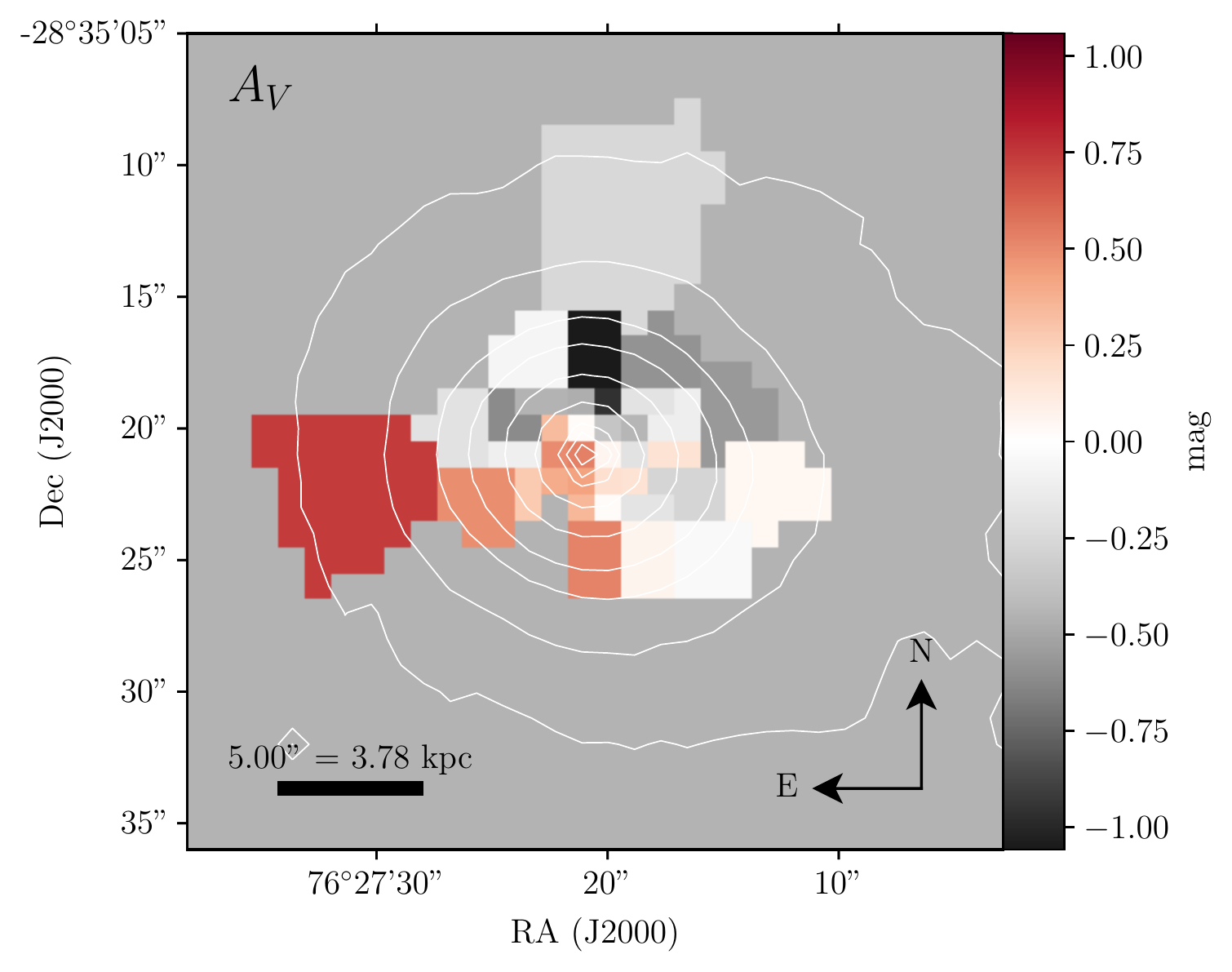}
	\caption{Total extinction in the $V$-band, $A_V$, calculated from the total \ha{} and \hb{} emission line fluxes. Negative values are probably a result of errors in the stellar continuum fit. The contours are as in Fig.~\ref{fig: stellar age map}.}
	\label{fig: A_V map}
\end{figure}

As shown in Fig.~\ref{fig: A_V map}, there is little extinction in the North-Western regions of the galaxy, whereas $A_V$ reaches values of $\sim 1 \,\rm mag$ in the South-East. 
Bins with $A_V < 0$ by more than $1\sigma$ are most likely due to systematic errors in the stellar continuum fit. 
There may be additional systematic errors due to our calculation of $A_V$ using the summed \ha{} and \hb{} fluxes from both broad and narrow components, which is problematic if the broad and narrow kinematic components have different intrinsic $A_V$ values.
Our assumption of the intrinsic \ha{}/\hb{} flux ratio of 2.86 may also not hold, which may be the case if the power source is AGN photoionisation or shocks~\citep[see Section~\ref{subsec: emission line ratios};][]{Dopita&Sutherland2003}.
Emission line fluxes were only corrected for extinction in bins in which $A_V > 0$.

\begin{table}
	\centering
	\caption{Total emission line fluxes, corrected for extinction. The quoted uncertainties represent $1\sigma$ errors derived from the formal uncertainties of the fits in each bin.} 
	\begin{tabular}{c c}
		\hline 
		\textbf{Line} & \textbf{Flux} ($10^{-15} \, \rm erg \, s^{-1} \, cm^{-2}$) \\
		\hline
		\forb{O}{ii}$\uplambda\uplambda 3726,9\text{\AA}$ & $ 5.2 \pm 0.7$ \\
		\hb{}$\uplambda4861\text{\AA}$ 	 & $ 7.6 \pm 0.7$ \\
		\forb{O}{iii}$\uplambda \uplambda 4959, 5007\,\text{\AA\AA}$ & $ 17.0 \pm 0.9 $ \\
		\forb{O}{i}$\uplambda 6300\text{\AA}$ & $4.5 \pm 0.3$ \\
		\ha{}$\uplambda6563\text{\AA}$ 	 & $ 20.7 \pm 0.5 $ \\
		\forb{N}{ii}$\uplambda \uplambda 6548, 83\text{\AA}$ 	 & $ 38.2 \pm 0.8 $ \\
		\forb{S}{ii}$\uplambda 6716\text{\AA}$ 	 & $ 18.3 \pm 0.5 $ \\
		\forb{S}{ii}$\uplambda 6731\text{\AA}$ 	 & $ 10.2 \pm 0.4 $ \\		
		\hline
	\end{tabular}
	\label{tab: emission line fluxes}
\end{table}

\subsection{Emission line ratios}\label{subsec: emission line ratios}
\begin{figure*}
	\includegraphics[width=1\linewidth]{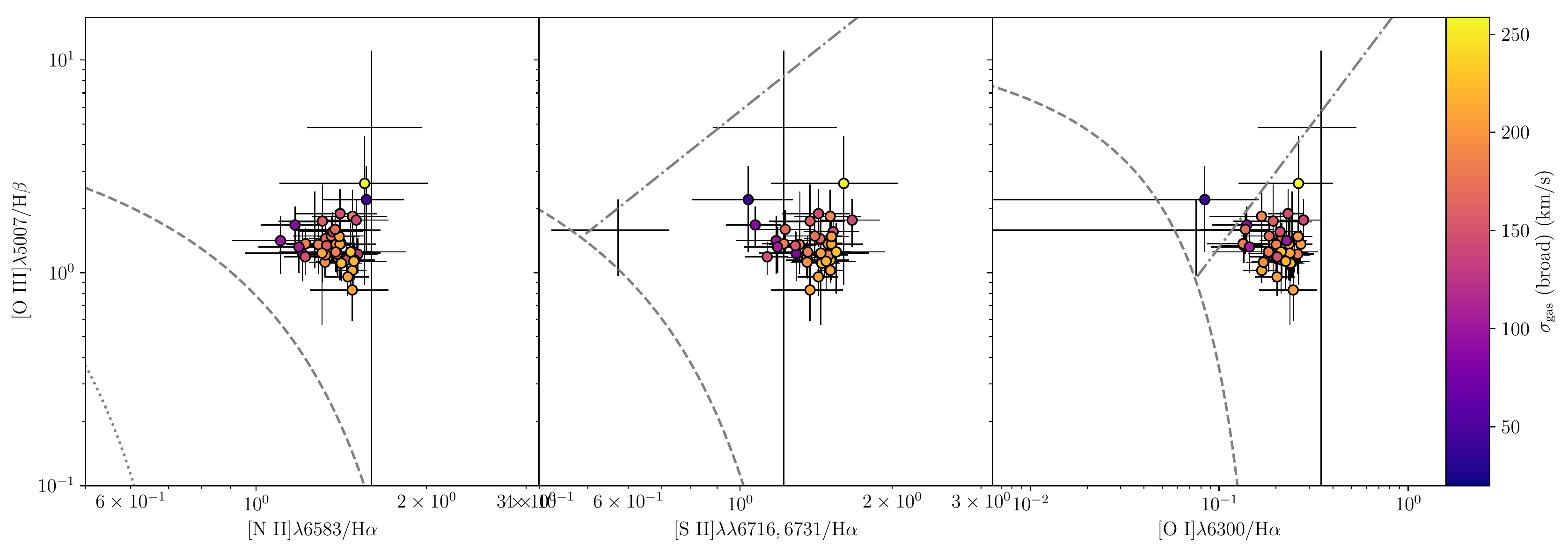}
	\caption{Optical diagnostic diagram \citep[ODD;][]{Baldwin1981,Veilleux&Osterbrock1987} where the line ratios have been computed from the \textit{total} (narrow plus broad) line fluxes in each bin, coloured by the velocity dispersion of the broad component.
	The grey dashed lines represent the maximum [O\,\textsc{iii}]/\hb{} ratio that can arise from star formation alone, derived from photoionisation models~\protect\citep{Kewley2001}.
	In the left panels, the dotted line is the equivalent empirical relation of \protect\citet{Kauffmann2003} which separates star-forming galaxies and AGN hosts. 
	In the middle and right panels, the dot-dashed lines of \protect\citet{Kewley2006} separate Seyfert-like (above) and LIER-like ratios (below the line).}
	\label{fig: BPT}
\end{figure*}

To investigate the source of the line emission, we used an optical diagnostic diagram~\citep[ODD;][]{Baldwin1981,Veilleux&Osterbrock1987} in which the ratios of pairs of emission lines are used to determine the excitation mechanism of the ionised gas. 
Due to the faintness of the bluer lines, we were unable to analyse the excitation mechanisms of each kinematic component separately; we therefore analysed the total fluxes summed from both broad and narrow components in each bin, as shown in Fig.~\ref{fig: BPT}, where each point has been coloured by the velocity dispersion of the broad component, which dominates the flux in most bins.
As is typical for a LERG such as \eso{}, the emission line ratios in all bins lie firmly in the Low-Ionisation Emission Region (LIER) quadrant, indicating ionisation dominated either by old stellar populations~\citep{Binette1994,Singh2013,Belfiore2016}, a low-luminosity AGN~\citep{Kauffmann2003b} or shocks~\citep{Allen2008,Dopita&Sutherland2017,Sutherland&Dopita2017}.

\subsection{Kinematics}\label{subsec: emission line kinematics}
To analyse the emission line kinematics, we carried out a separate line fitting procedure for the \ha{} and \forb{N}{ii} lines in the R7000 cubes in order to take advantage of the higher spectral resolution.
To create the emission line-only spectra for each bin in the R7000 data cube, we used a cubic spline technique to spectrally interpolate the stellar continuum from the \ppxf{} fit to the spectral resolution of the R7000 data cube. The interpolated stellar continua were then subtracted from the spectrum in each bin. The fit was then carried out using the same method described in Section~\ref{sec: emission line analysis; subsec: emission line fitting}.

\begin{figure*}
	\centering
	\includegraphics[width=0.65\linewidth]{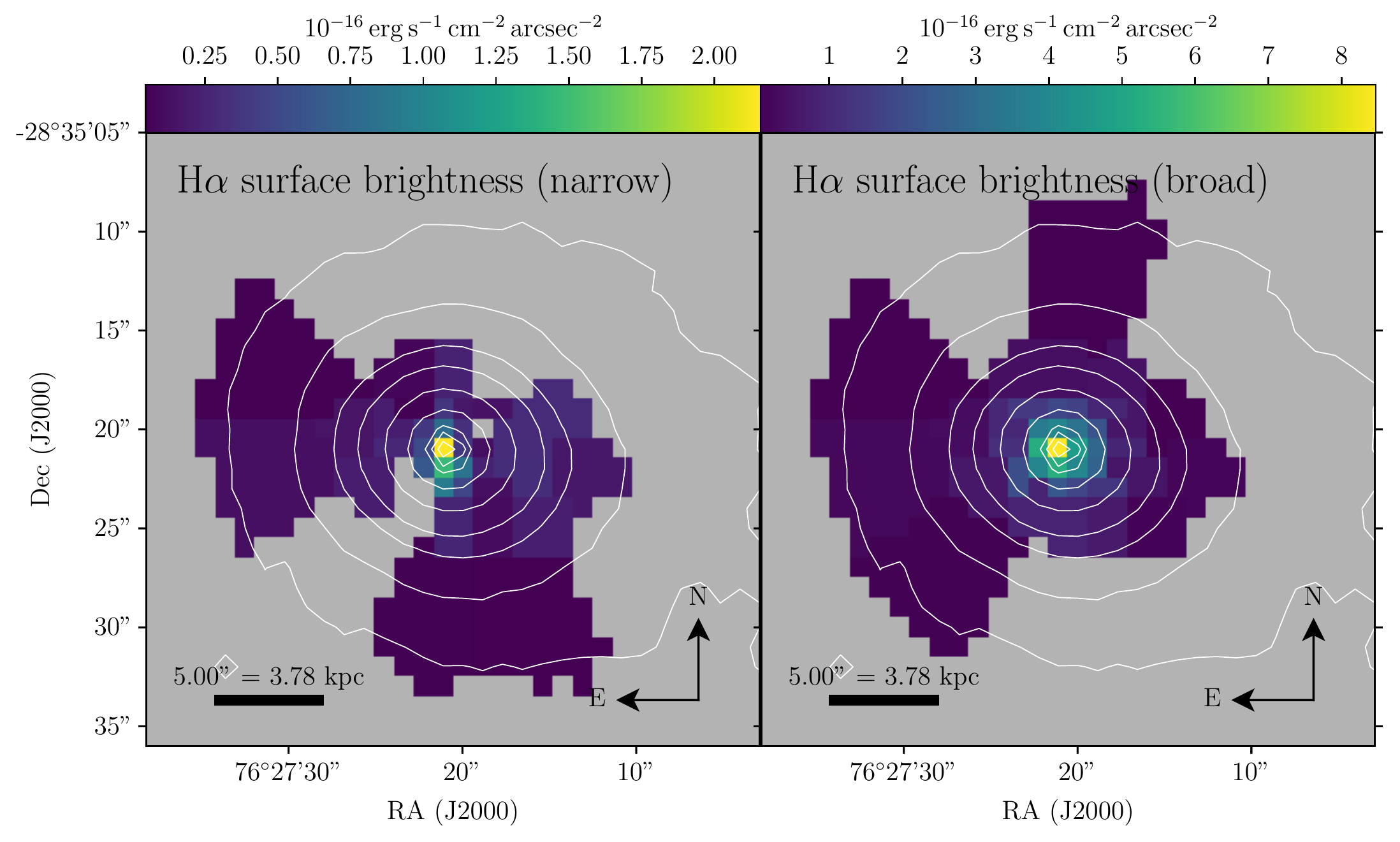}
	\caption{\ha{} surface brightness of the narrow (left) and broad (right) kinematic components. The contours are as in Fig.~\ref{fig: stellar age map}.}
	\label{fig: Halpha surface brightness}
\end{figure*}

In most bins, one broad and one narrow component were required to adequately describe the line profiles.
In bins where there were multiple fits with differing kinematics but similar reduced-$\chi^2$ values, the fit that would produce the smoothest variation in kinematics across neighbouring bins was chosen.
The surface brightness of the \ha{} line in each component is shown in Fig.~\ref{fig: Halpha surface brightness}; the broad component is significantly brighter than the narrow component, which appears to increase slightly to the West.

\begin{figure*}
	\includegraphics[width=0.9\linewidth]{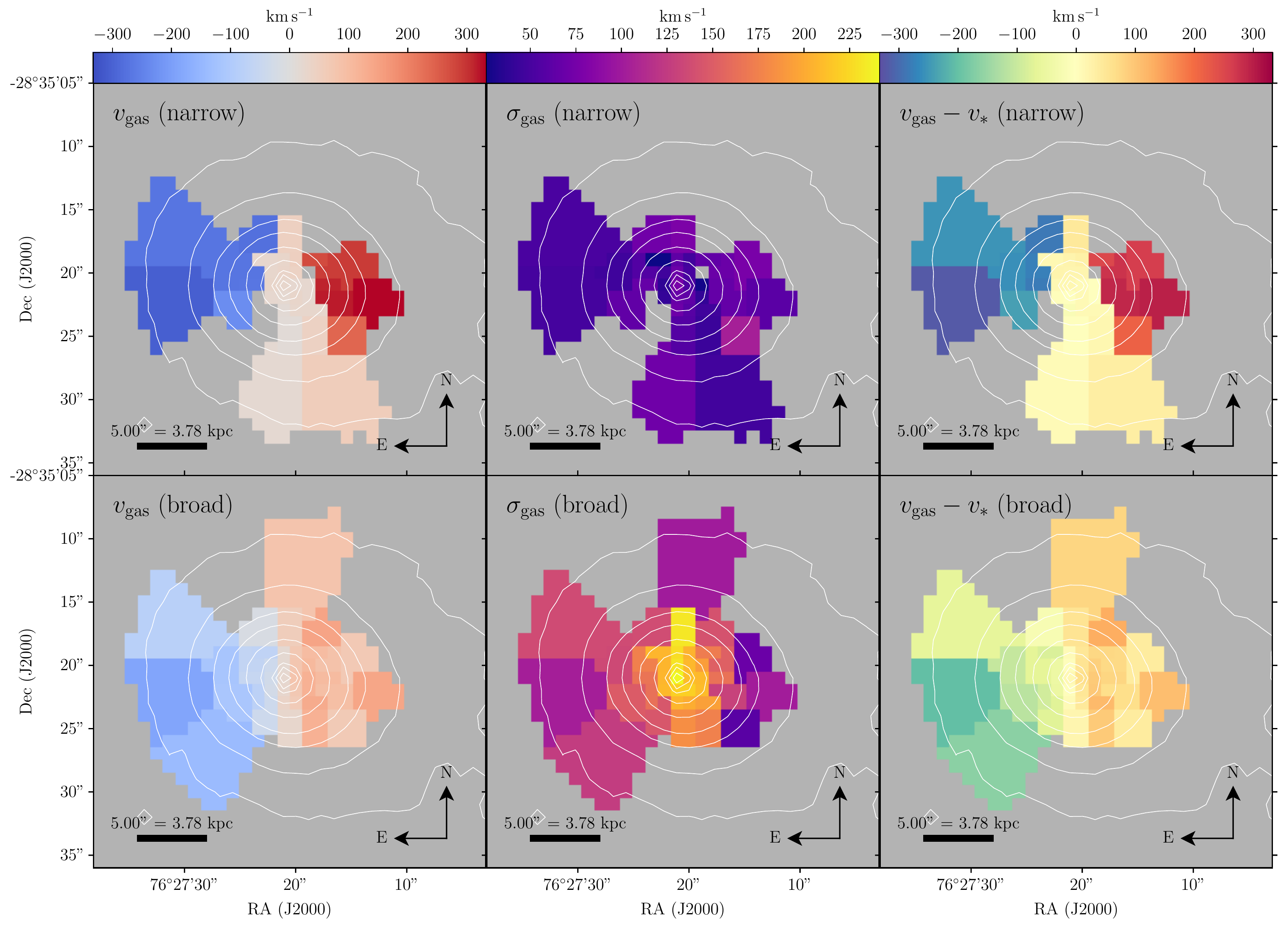}
	\caption{Ionised gas radial velocity, relative to systemic (left), gas velocity dispersion (Gaussian $\sigma$, minus instrumental dispersion, right) and gas velocity relative to the stellar velocity (centre) of the narrow (top row) and broad (bottom row) emission line components in each bin. The contours are as in Fig.~\ref{fig: stellar age map}.}
	\label{fig: Halpha kinematics}
\end{figure*}

The radial velocity and velocity dispersion of each component is shown in Fig.~\ref{fig: Halpha kinematics}. Also shown is the difference between the stellar and gas velocities for each component.
Ordered rotation is apparent in both the broad and narrow components, although the rotation curve of the narrow component is much steeper than that of the broad component. The rotation curves of both components are also much steeper than that of the stars: whilst the stellar velocities peak at $35 \,\rm km\,s^{-1}$, the broad and narrow components reach maximum radial velocities of approximately $100 \,\rm km\,s^{-1}$ and $320 \,\rm km\,s^{-1}$ respectively.
The velocity dispersions of the two components also differ vastly: whilst that of the broad component rises smoothly towards the centre, peaking at approximately $225 \,\rm km\,s^{-1}$, the width of the narrow component remains relatively constant at approximately $60 \,\rm km\,s^{-1}$ across the entire galaxy. 

The starkly different rotation curves of the broad and narrow components, combined with their respective widths, suggests that each component traces a distinct disk.
Similar asymmetric line profiles observed in so-called ``red geyser'' galaxies have recently been attributed to AGN-driven biconical outflows~\citep{Roy2021a}. Indeed, \eso{} shares many properties of red geyser galaxies, intermediate-mass ($M_* \sim 10^{10.5}\,\rm M_\odot$) spheroidal galaxies characterised by low star formation rates and misaligned gas and stellar kinematics~\citep{Cheung2016}.
However, were the asymmetric line profiles in \eso{} a result of biconical outflows, this would imply that the conical structures are perpendicular to the kpc-scale jets (see Fig.~\ref{fig: FoV (small)}), which would be difficult to explain were the outflows driven by the central AGN.
We therefore favour the multiple disk interpretation, although our ability to carry out detailed kinematic modelling is hampered by low S/N and the correspondingly large bins in the outermost regions.

\subsection{Star formation rate}\label{subsec: star formation rate}
The star formation rate (SFR) was first estimated using the SFH obtained from our stellar population analysis (Section~\ref{sec: Stellar population analysis}). As shown in Fig.~\ref{fig: stellar age map}, stellar populations less than 100 Myr old account for approximately $10^{7.6}\,\rm M_\odot$, corresponding to a mean SFR of $0.4 \,\rm M_\odot \, yr^{-1}$.

We also estimated the SFR using the \ha{} calibration of \citet{Calzetti2013}, ${\rm SFR(H\upalpha)} = 5.5 \times 10^{42} L_{\rm H\upalpha}\,\rm M_\odot\,yr^{-1}$, assuming a Kroupa IMF and a constant SFR over a timescale $\tau \geq 6\,\rm Myr$.
The total extinction-corrected \ha{} luminosity, comprising both broad and narrow components, yields ${\rm SFR(H\upalpha)} = 0.39 \pm 0.01\,\rm M_\odot\,yr^{-1}$.
As shown in Fig.~\ref{fig: BPT}, the emission line ratios are LIER-like, indicating the \ha{} emission is dominated by processes other than star formation; this SFR estimate therefore represents an upper limit, although we note that it is consistent with the SFR estimate based on our \ppxf{} analysis.
The corresponding specific SFR, i.e., the star formation rate per unit stellar mass, of \eso{} is approximately $\rm sSFR \sim 0.001 \,\rm Gyr^{-1}$.



\subsection{Eddington ratio}\label{subsec: Eddington ratio}

Here we compute the Eddington ratio
\begin{equation}
\lambda_{\rm Edd} = \frac{L_{\rm bol}}{L_{\rm Edd}}
\label{eq: Eddington ratio}
\end{equation}
were $L_{\rm bol}$ is the bolometric luminosity of the AGN and $L_{\rm Edd}$ is the Eddington ratio.
To compute $L_{\rm bol}$ we use the relation of \citet{Heckman2004}
\begin{equation}
L_{\rm bol} = 3500 L_{\rm [O\,III]} \,\rm erg\,s^{-1}
\end{equation}
where $L_{\rm [O\,III]}$ is the luminosity of the \forb{O}{iii}$\uplambda 5007\text{\AA}$ line. Using the integrated fluxes shown in Table~\ref{tab: emission line fluxes} we find $L_{\rm [O\,III]} = (4.3 \pm 0.2) \times 10^{40}\,\rm erg \, s^{-1}$, corresponding to $L_{\rm bol} = (1.50 \pm 0.08) \times 10^{44}\,\rm erg\,s^{-1}$.

The Eddington luminosity is given by
\begin{equation}
L_{\rm Edd} = 1.3 \times 10^{38} \left(\frac{M_{\rm BH}}{\rm M_\odot}\right) \,\rm erg\,s^{-1}
\end{equation}

To estimate the supermassive black hole mass $M_{\rm BH}$ we use the $M_{\rm BH} - \sigma_*$ relation~\citep{Ferrarese2000,Gebhardt2000,Kormendy&Ho2013}
\begin{equation}
\frac{M_{\rm BH}}{10^9 \,{\rm M_{\odot}}} = \left(0.310^{+0.037}_{-0.033}\right) \left(\frac{\sigma_*}{200 {\rm \, km\,s^{-1}}}\right)^{4.38 \pm 0.29}
\end{equation}
where $\sigma_*$ is the stellar velocity dispersion within $1R_e$.
To compute $\sigma_*$ we used \ppxf{} to fit the stellar continuum to the spectrum extracted from an aperture with radius $1R_e = 12.56$'' and an axis ratio of 1.25~\citep{Lauberts&Valentijn1989}. We find $\sigma_* = 293 \,\rm km\,s^{-1}$, which gives $M_{\rm BH} = 1.7^{+0.4}_{-0.3} \times 10^9 \,\rm M_\odot$ and a corresponding Eddington luminosity $L_{\rm Edd} = 2.2^{+0.5}_{-0.4} \times 10^{47}\,\rm erg\,s^{-1}$. 

Substituting these values into Eqn.~\ref{eq: Eddington ratio} gives $\lambda_{\rm Edd} = (6 \pm 2) \times 10^{-4}$, suggesting radiatively inefficient accretion as is characteristic of LERGs~\citep{BestHeckman2012}.
Both the Eddington ratio and BH mass of \eso{} are also characteristic of the sample of low-excitation GRGs studied by \citet{Dabhade2020b}.

The mass accretion rate required to power the AGN can be estimated using
\begin{equation}
\dot{M} = \frac{L_{\rm bol}}{c^2 \eta}
\end{equation}
where $\eta \approx 0.1$ is the efficiency~\citep{Riffel2008}; adopting $L_{\rm bol} \sim 10^{44} \rm \, erg\,s^{-1}$ gives $\dot{M} \sim 10^{-2} \,\rm M_\odot \, yr^{-1}$.

\section{Na D absorption line analysis}\label{sec: Na D absorption line analysis}

As illustrated in Fig.~\ref{fig: example ppxf fits}, the best-fit stellar continua in most bins significantly under-estimates the strength of the Na\,D absorption, suggesting this feature is partially interstellar in origin. With an ionisation potential of $5.1\,\rm eV$, interstellar Na\,D absorption is a tracer of neutral gas.

To further investigate these residuals, we normalised the spectra in each bin of the R7000 cube by the stellar continuum as follows.
In each bin, the stellar continuum from the \ppxf{} fit was spectrally interpolated using a cubic spline technique to the spectral resolution of the R7000 data cube. The spectrum was then divided by the interpolated stellar continuum.

\begin{figure*}
	\includegraphics[width=0.85\linewidth]{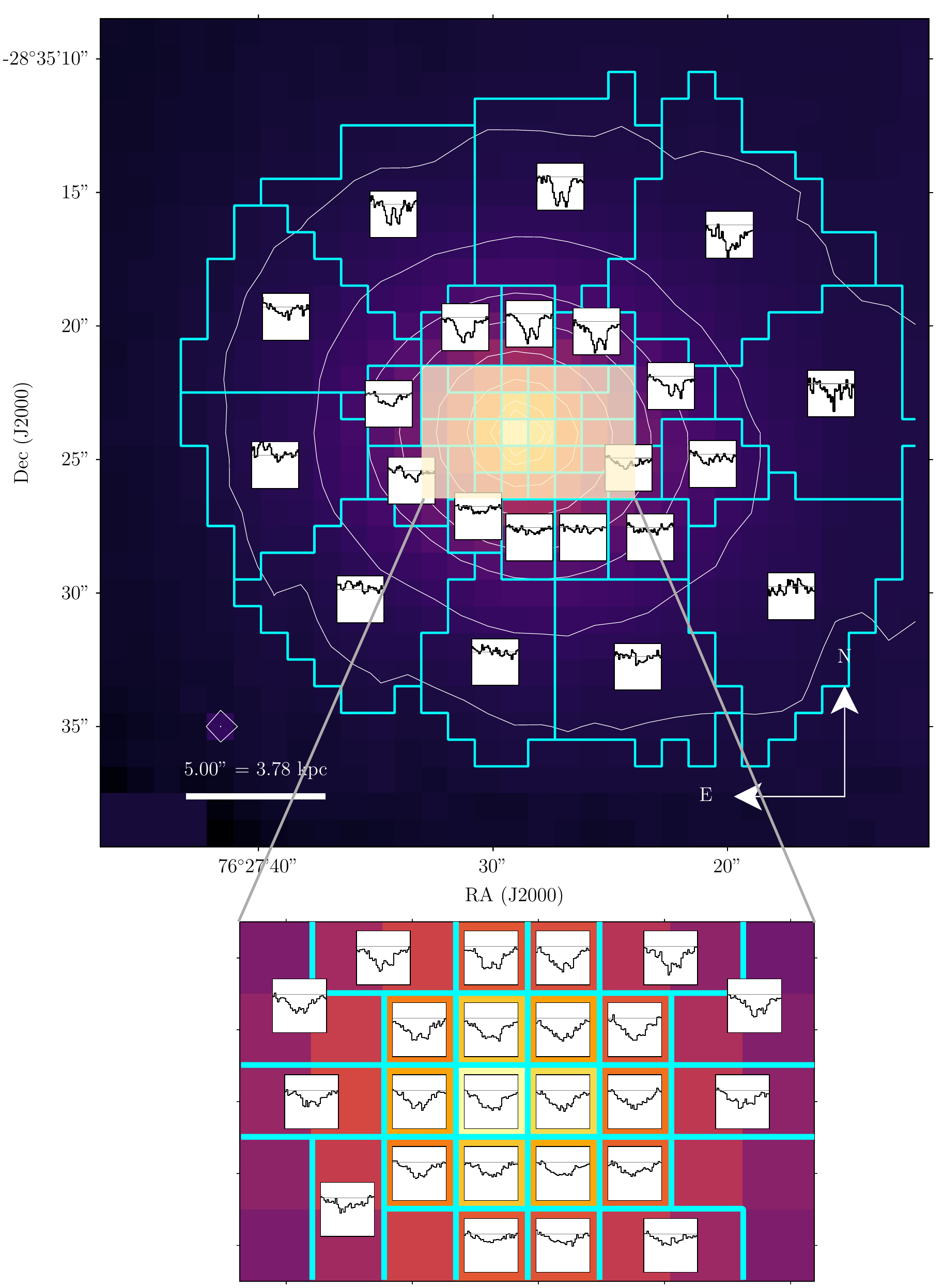}
	\caption{The residual Na D absorption profile in each bin (denoted by the cyan outlines), overlaid on the $V$-band continuum. The lower plot shows an enlarged view of the central region. Each inset set of axes has the same scaling on the $x$- and $y$-axes. The grey horizontal line denotes $F_\uplambda/F_{*,\uplambda} = 1.0$. The contours are as in Fig.~\ref{fig: stellar age map}.}
	\label{fig: Na D absorption profiles}
\end{figure*}

The residual absorption profiles in each bin are shown in Fig.~\ref{fig: Na D absorption profiles}.
A variety of profile shapes are present: in some bins, the lines of the doublet are clearly visible, whereas in others they are unresolved. In many bins, the profiles are noticeably asymmetric, exhibiting clear broad and narrow kinematic components.
The overall absorption strength increases smoothly from South-East to North-West; interestingly, the deepest Na\,D residuals are found in regions with $A_V \sim 0$ (Fig.~\ref{fig: A_V map}), whereas the regions with greater extinction exhibit little to no interstellar Na\,D absorption. In most galaxies, interstellar Na\,D absorption is correlated with $A_V$ due to dust in the absorbing gas~\citep[e.g.,][]{Rupke2021,Cheung2016}. 

We used the analytical expressions for Na\,D absorption profiles given by \citet{Rupke2005a} to fit the residual Na\,D absorption features in each bin, where we assumed a Maxwellian velocity distribution for the absorbing gas, and a covering fraction $C_f$ that does not vary with wavelength. 
For each kinematic component, the two lines in the doublet were modelled assuming the ``completely overlapping'' case, such that the intensity $I(\uplambda)$ of the absorption, assuming an incident continuum level of unity, is given by
\begin{equation}
I(\uplambda) = 1 - C_f + C_f e^{-\tau_b(\uplambda) - \tau_r(\uplambda)}
\end{equation}
where $\tau_b(\uplambda)$ and $\tau_r(\uplambda)$ are the optical depths of the blue and red lines in the doublet~\citep[at $\uplambda_b = 5889.9$\,\AA{} and $\uplambda_r = 5895.9$\,\AA{} in air respectively;][]{Morton1991}.
Under the assumption of a Maxwellian velocity distribution, the optical depth of line $i$ centred at wavelength $\uplambda_i$ is a Gaussian of the form
\begin{equation}
\tau_i(\uplambda) = \tau_{i, 0} \exp\left[\frac{- (\uplambda - \uplambda_i)^2}{(\uplambda_i b / c)^2}\right]
\end{equation}
were $\tau_{0, i}$ is the ``central'' optical depth at $\uplambda = \uplambda_i$, $c$ is the speed of light and $b$ is the Doppler line width, related to the Gaussian $\sigma$ via $\sigma = b / \sqrt{2} c$. We assumed $\tau_{b, 0} = 2\tau_{r, 0}$~\citep{Morton1991}.

Two kinematic components were fitted to the absorption profiles, combining each component $I_1(\uplambda)$ and $I_2(\uplambda)$ (each consisting of the two lines in the doublet) assuming the ``partial overlap'' case of \citet{Rupke2005a}, in which case the residual absorption profile is given by
\begin{equation}
I(\uplambda) = I_1(\uplambda) I_2(\uplambda).
\end{equation}

At widths $b \gtrsim 300\,\rm km\,s^{-1}$ the two lines in the doublet become unresolved, and the line shape approaches a Gaussian or Voigt-like profile. In this regime, it becomes difficult to estimate both $C_f$ and $\tau_0$: within the measurement uncertainties of our data, an absorber with large covering fraction and low optical depth may become indistinguishable from one with low covering fraction and high optical depth.
In these cases, $C_f$ and $\tau_0$ become strongly degenerate, and the formal uncertainties provided by simple least-squares minimisation techniques are inaccurate as they do not reflect this anticorrelation. 

We instead opted to use a Bayesian parameter estimation technique to fit the line profiles in each bin, as these approaches numerically estimate the posterior probability distribution functions (PDFs) of the model parameters given the data, enabling correlations between parameters to be easily visualised and interpreted.
We used \textit{nested sampling}~\citep{Skilling2004,Skilling2006} to compute the PDFs of the model parameters $\mathbf{\Theta}$ given the data $\mathbf{D}$ and the model $M$. Other Bayesian parameter estimation techniques, such as Monte-Carlo Markov Chain and dynamical nested sampling, have also been used to fit Na\,D absorption profiles in the past~\citep[e.g.,][]{Sato2009,Roy2021b}.

The posterior $\mathbf{P}\left(\mathbf{\Theta} | \mathbf{D}, M\right)$ is given by Bayes' theorem,

\begin{equation}
\mathbf{P}\left(\mathbf{\Theta} | \mathbf{D}, M\right) = \frac{\mathbf{P}\left(\mathbf{D} | \mathbf{\Theta}, M\right) \mathbf{P}\left(\mathbf{\Theta} | M\right)}{\mathbf{P}\left(\mathbf{D} | M\right)}
\label{eq: Bayes theorem}
\end{equation}

where $\mathbf{P}\left(\mathbf{D} | \mathbf{\Theta}, M\right)$ is the \textit{likelihood} of observing the data given the parameters and the model, and $\mathbf{P}\left(\mathbf{\Theta} | M\right)$ is the \textit{prior} probability distribution of the model parameters.
$\mathbf{P}\left(\mathbf{D} | M\right)$ is the \textit{Bayesian evidence}, or the likelihood of the data given the model, which is the integral of the likelihood and the prior over all parameter space:
	
\begin{equation}
\mathbf{P}\left(\mathbf{D} | M\right) = \int_{\Omega_{\mathbf{\Theta}}} \mathbf{P}\left(\mathbf{D} | \mathbf{\Theta}, M\right) \mathbf{P}\left(\mathbf{\Theta} | M\right) d\mathbf{\Theta}
\label{eq: evidence integral}
\end{equation}

%
%

Nested sampling produces an estimate of $\mathbf{P}\left(\mathbf{D} | M\right)$ by approximating the integral in Eqn.~\ref{eq: evidence integral} via computing ``iso-likelihood'' surfaces in parameter space $\Omega_{\mathbf{\Theta}}$. 
The likelihood and the prior at the sampled points in parameter space are also evaluated as part of this computation, enabling the posterior to be sampled as per Eqn.~\ref{eq: Bayes theorem}. In this way, nested sampling can be used to approximate the posterior PDFs of model parameters.

To estimate posterior probability distributions and the Bayesian evidence for our models, we used the nested sampling Monte Carlo algorithm \textsc{MLFriends}~\citep{Buchner2014,Buchner2019} implemented in the \textsc{python} package \textsc{Ultranest}\footnote{\url{https://johannesbuchner.github.io/UltraNest/}}~\citep{Buchner2021}.

In each bin, we estimated the posterior PDF for absorption profile models comprising both one and two kinematic components, where the best-fit number of components in each bin was determined by eye.
We did not convolve the model profiles with the instrumental resolution as this greatly increased the computational time, and is expected to have only a modest impact on the results~\citep{Rupke2005a}.
The reported best-fit parameter values, and corresponding upper and lower error estimates, for each component were computed from the quartiles of the samples in the marginalised PDFs.

We adopted uniform priors for our model parameters.
The covering fraction was limited to $C_f \in [0, 1]$, and the optical depth was limited to $\tau_0 \in [0, 5]$ because the lines saturate at optical depths $\tau_0 \geq 5$. 
The radial velocity was constrained to $v \in [-100, 500]\rm \, km\,s^{-1}$ relative to systemic. 
For two-component fits, we fitted a narrow and a broad component by defining priors $b_{\rm narrow} \in [10, 200]\rm \, km\,s^{-1}$ and $b_{\rm broad} \in [200, 800]\rm \, km\,s^{-1}$. For single component fits we set $b \in [10, 800]\rm \, km\,s^{-1}$.

Examples of the best-fit line profiles and corresponding corner plots for two bins are given in Appendix~\ref{sec: appendix: example NaD fits}. 
The marginalised posterior PDFs for both $C_f$ and $\tau_0$ tend to have long tails extending to large values, resulting in significantly asymmetric upper and lower error estimates. Strong correlations between $C_f$ and $\tau_0$ are also present in most bins. 
In bins where two components were required, the PDFs corresponding to the parameters of the narrow component tend to be well-defined and quasi-Gaussian (Fig.~\ref{fig: appendix: example NaD fit (bin 0)}) whereas the PDFs for the broad component parameters tend to be irregular and even multi-modal (Fig.~\ref{fig: appendix: example NaD fit (bin 41)}), reflecting the difficulties in constraining model parameters when the two lines become blended.

\begin{figure*}
	\includegraphics[width=0.9\linewidth]{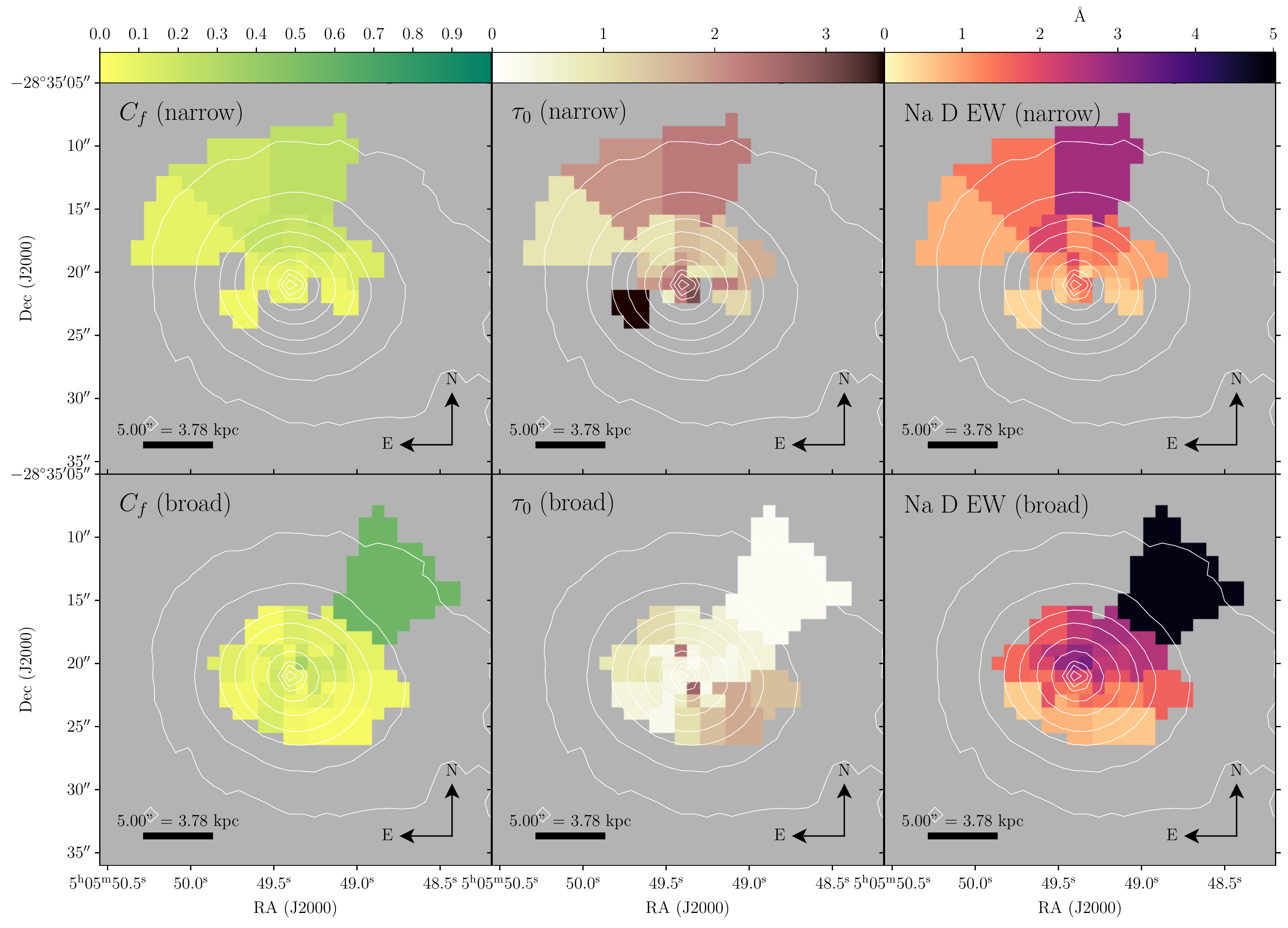}
	\caption{Covering fraction $C_f$ (left), central optical depth $\tau_0$ of the Na\,D red line (centre) and EW (right) of the best-fit narrow (top row) and broad (bottom) kinematic components. The contours are as in Fig.~\ref{fig: stellar age map}.}
	\label{fig: NaD parameters}
\end{figure*}

Fig.~\ref{fig: NaD parameters} shows the 50th percentile $\tau_0$ and $C_f$ values in each bin. 
The narrow component is optically thick, with $\tau_0 \geq 1$ in most bins, but with low covering fractions, suggesting this component traces dense clumps.
Meanwhile, the broad component is optically thin ($\tau_0 \lesssim 1$) with similarly low covering fractions.

\begin{figure*}
	\includegraphics[width=0.9\linewidth]{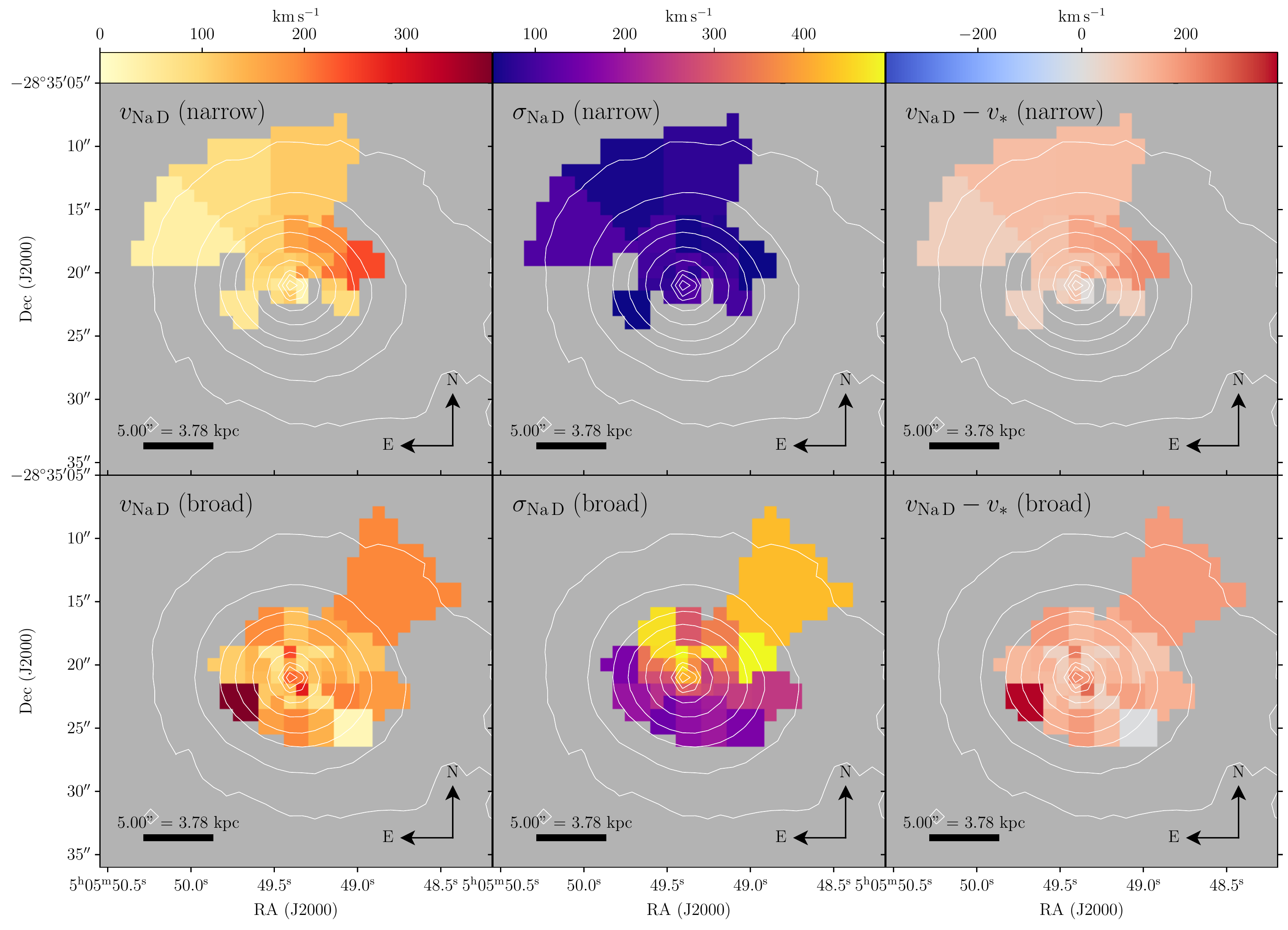}
	\caption{The radial velocity (left) and velocity dispersion (centre) of the narrow (top row) and broad (bottom row) components of the residual Na\,D absorption. The right column shows the difference between the radial velocity of the Na\,D and the best-fit stellar population. The contours are as in Fig.~\ref{fig: stellar age map}.}
	\label{fig: Na D kinematics}
\end{figure*}

As shown in Fig.~\ref{fig: Na D kinematics}, both narrow and broad components are redshifted by up to $+250 \,\rm km\,s^{-1}$ with respect to systemic velocity.
Importantly, because the Na\,D lines are viewed in absorption against the stellar continuum, this indicates the absorption traces \textit{inflowing} material.
The narrow component is characterised by velocity dispersions of 60--100\,$\rm km\,s^{-1}$ and exhibits a velocity shear similar to that of the ionised gas, suggesting this gas traces the same rotating structure (see Fig.~\ref{fig: Halpha kinematics}). In contrast, the broad component shows no clear signs of rotation, and is very broad, with $\sigma \geq 500 \,\rm km\,s^{-1}$ in some bins.

Also shown in Fig.~\ref{fig: Na D kinematics} is the difference between the stellar and Na\,D radial velocities. 
To determine whether the absorption is sufficiently redshifted to represent an inflow, we adopted the the $2\sigma_{v_{\rm Na\,D}}$ criterion of \citet{Krug2010} where $\sigma_{v_{\rm Na\,D}}$ is the lower error on the LOS Na\,D velocity.
For the narrow component, the interstellar Na\,D meets this requirement in most bins. The offset of the broad component does not exceed the $2\sigma$ criterion in roughly half of the bins, although this is likely a result of the larger uncertainty on the fitted velocity due to the broadness of the profile.

Fig.~\ref{fig: NaD parameters} shows the rest-frame equivalent width (EW) calculated from the best-fit $I(\uplambda)$ in each bin, using
\begin{equation}
{\rm EW_{tot}} = \frac{1}{1 + z} \sum_{i} \left\lbrace \int_{0}^{\infty}\left[1 - I_{b, i}(\uplambda)\right] d\uplambda + \int_{0}^{\infty}\left[1 - I_{r, i}(\uplambda)\right] d\uplambda \right\rbrace
\end{equation}
where $I_{b, i}(\uplambda) = 1 - C_{f, i} + C_{f, i}e^{-\tau_{b,i}(\lambda)}$ and $I_{r, i}(\uplambda) = 1 - C_{f, i} + C_{f, i}e^{-\tau_{r,i}(\lambda)}$ are the red and blue components of the doublet in each kinematic component $i$. 
The EW ranges from 1--5\,\AA{} and increases smoothly from South-East to North-West, suggesting that the gas is inflowing from the North.
Interestingly, the EWs are significantly higher than those observed in a sample of local active and inactive galaxies with Na\,D inflows reported by \citet{Roberts-Borsani&Saintonge2019}, who detected EWs of $\lesssim 1.3$\,\AA{}, and our measured EWs are among the highest of those detected in red geyser galaxies~\citep{Roy2021b}.
The EWs are more similar to those observed in the sample of ultra-luminous infrared galaxies (ULIRGs) by \citet{Rupke2005b} (0--10\,\AA), all of which host outflows.

In addition to interstellar absorption, stellar template mismatch has been observed to cause significant Na\,D residuals in early-type galaxies. 
The mismatch is thought to result from elevated Na abundances arising from non-solar abundance scalings, and leaves broad, residual absorption features centred at the systemic stellar velocity~\citep{Jeong2013,Park2015}.
In \eso{}, the narrow component is clearly interstellar in origin, as evidenced by its low velocity dispersion ($\sigma_{\rm Na\,D} \lesssim 100\,\rm km\,s^{-1}$) compared to that of the stellar component ($\sigma_* \gtrsim 250 \rm \,km\,s^{-1}$; see Fig.~\ref{fig: stellar kinematics}).
The broad component, on the other hand, has velocity dispersions similar to that of the stars (Fig.~\ref{fig: stellar kinematics}); however, both the velocity dispersion and the EW of the broad component increase markedly from South to North. 
In fact, inspection of Fig.~\ref{fig: Na D absorption profiles} reveals negligible residual absorption -- broad or narrow -- in the Southern regions of the galaxy, indicating the broad component is predominantly interstellar in origin, although some contamination from residual stellar absorption may be present.
We therefore repeated our analysis, substituting a stellar component, that was fixed to the stellar velocity and velocity dispersion, for the broad interstellar component. As detailed in Appendix~\ref{sec: appendix: residual stellar Na D absorption}, the resulting amplitude of the stellar component varies significantly across the galaxy, confirming that the broad component is predominantly interstellar.

\subsection{Mass inflow rate}
To estimate the mass inflow rate, we used the method described by \citet{Rupke2005a}.
We note that this method was originally developed to estimate mass outflow rates, although the same technique was applied to inflows by \citet{Krug2010}, with some caveats which we discuss here.

The inflow was modelled as a mass-conserving flow, such that the mass inflow rate $\dot{M}$ and velocity $v$ are independent of radius. We used the ``thin shell'' approximation, in which the Na\,D absorption is assumed to arise from a thin shell of material at a distance $r$ from the galactic centre. Under this assumption, the mass inflow rate averaged over the lifetime of the wind is given by 
\begin{equation}
\dot{M} = \Omega \mu m_p N({\rm H}) v r = 4 \uppi C_\Omega C_f \mu m_p N({\rm H}) v r,
\end{equation}
where $\mu m_p$ is the average particle mass (where $m_p$ is the proton mass and $\mu = 1.44$), and $N({\rm H})$ is the \hi{} column density.
$\Omega =  4 \uppi C_\Omega C_f$ is the solid angle subtended by the inflow viewed from the galactic centre, which includes both the large-scale covering fraction $C_\Omega$ as well as that representing the clumpiness of the wind, $C_f$.

For our radius we chose $r = 8\,\rm kpc$ as it represents the maximum radius at which Na\,D absorption is detected, and to calculate the large-scale covering fraction of each bin we used $C_\Omega = A_{\rm bin} / 4 \uppi r^2$ where $A_{\rm bin}$ is the area of the bin projected onto the plane of the sky. 

To calculate the hydrogen column density $N({\rm H})$, we first used the expression given by \citet{Spitzer1978} to compute the Na\,\textsc{i} column density,
\begin{equation}
N({\rm Na\,\textsc{i}}) = \frac{\tau_{b, 0} b}{1.497 \times 10^{-15} \uplambda_b f_b}\,\rm cm^{-2}
\label{eq: N(Na I)}
\end{equation}
where $f_b = 0.3180$ is the oscillator strength~\citep{Morton1991}. $N({\rm H})$ is then 
\begin{equation}
N({\rm H}) = \frac{N({\rm Na\,\textsc{i}})}{\left(1 - y\right)10^{(a + b_{\rm dep})}},
\label{eq: N(H)}
\end{equation}
where $y$ is the ionisation fraction, $a$ is the Na abundance and $b_{\rm dep}$ is the depletion factor of Na onto dust.
Because we could not constrain any of these values from our observations, we assumed $y = 0.9$, solar Na abundance~\citep[$a = -5.69$;][]{Savage&Sembach1996} and Galactic dust depletion~\citep[$b_{\rm dep} = -0.95$;][]{Savage&Sembach1996}. 
$N(\rm H)$ ranges from $10^{20} - 10^{22}\,\rm cm^{-2}$, similar to those measured in samples of radio galaxies with redshifted \hi{} absorption by \citet{vanGorkom1989}.


The mass inflow rate in each bin, comprising both broad and narrow absorption components, is then
\begin{equation}
\dot{M}_i = \Omega_i \mu m_p \left[N_{i,\rm broad}({\rm H}) v_{i,\rm broad} + N_{i,\rm narrow}({\rm H}) v_{i,\rm narrow} \right]r.
\label{eq: Mdot}
\end{equation}

To derive estimates for $N(\rm Na\,\textsc{i})$, $N(\rm H)$ and $\dot{M}$ in each bin, we evaluated Eqns.~\ref{eq: N(Na I)}, \ref{eq: N(H)} and \ref{eq: Mdot} for each sample of the posterior PDF generated by \textsc{Ultranest}. We then estimated central values and errors by computing 16th, 50th and 84th percentiles for the resulting distribution of each parameter.
The total mass inflow rates for the narrow and broad components are $\dot{M}_{\rm narrow} = 1.8^{+0.8}_{-0.6} \,\rm M_\odot \, yr^{-1}$ and $\dot{M}_{\rm broad} = 0.4^{+0.4}_{-0.2} \,\rm M_\odot \, yr^{-1}$, corresponding to a total mass inflow rate $\dot{M} \sim 1 - 3 \,\rm M_\odot \, yr^{-1}$.


There are many assumptions associated with this calculation.
We assumed a ``thin shell'' geometry such that all of the material is located in a thin shell with a radius of 8 kpc, whereas the increase in radial velocity of the narrow component towards the West (Fig.~\ref{fig: Na D kinematics}) may represent gas at smaller radii that has accelerated as it has fallen inwards. 
Additionally, inflow velocities are likely underestimated as the velocities are projected along the line-of-sight; the true velocity in most bins may be substantially higher.
There are also many assumptions in our calculation of $N({\rm H})$, including the Na abundance and ionisation fraction: in particular, because the inflowing gas is likely to be metal-poor (see Section~\ref{subsec: the origin of the inflow}), our $N({\rm H})$ values are probably underestimated. 
We estimate these factors to impact the inflow rate by an order of magnitude.

\section{Discussion}

\subsection{The young stellar population}

Our \ppxf{} analysis revealed a young ($\lesssim 10 \,\rm Myr$) stellar population in the Northern regions of \eso{} with an estimated mass of $10^{7.6}\,\rm M_\odot$. Although low levels of ongoing star formation are common in fast rotator ETGs, this is highly unusual in a slow rotator such as \eso{}~\citep{Shapiro2010}.

Simulations predict that jets can trigger star formation by facilitating instabilities and cloud collapse~\citep{Gaibler2012,Fragile2004,Fragile2005,Fragile2017}.
We can safely rule out this scenario in \eso{}; jets tend to inject energy and momentum spherically on kpc scales, thereby leading to a global elevation in the SFR~\citep{Mukherjee2016}, in contrast to the relatively localised region of young stars in \eso{}.

It is more likely that the young stellar population formed in-situ after a merger or accretion event.
Young stellar populations are found in 15--25\,per\,cent of radio galaxies~\citep{Tadhunter2011}, which has been attributed to gas-rich mergers simultaneously triggering star formation and jet activity.
Indeed, the estimated mass inflow rate estimated from our Na\,D observations ($\dot{M} \sim 1 - 3 \,\rm M_\odot \, yr^{-1}$) is sufficient to sustain the observed SFR ($0.4\,\rm M_\odot \, yr^{-1}$; see Section~\ref{subsec: star formation rate}).
Although \eso{} shows no obvious morphological disturbances evident of a merger, minor gas-rich mergers can provide fuel for star formation without producing a detectable morphological perturbation~\citep{Huang&Gu2009,Jaffe2014}.

Alternatively, \eso{} may have acquired the already-formed young stellar populations via a merger. 
With our current observations, there is unfortunately no way to determine whether the stars were formed in-situ or acquired externally via a merger.

\subsection{An inflow of neutral gas}\label{subsec: the origin of the inflow}

As discussed in Section~\ref{subsec: emission line kinematics}, the ionised gas in \eso{} exhibits dynamically distinct narrow ($\sigma \lesssim 100\,\rm km\,s^{-1},\, v_{\rm rot} \approx 300 \,\rm km\,s^{-1}$) and broad ($\sigma \gtrsim 150\,\rm km\,s^{-1},\,v_{\rm rot} \approx 100 \,\rm km\,s^{-1}$) kinematic components. The ionised gas is clearly not in equilibrium with the stellar component, which has a maximum LOS rotational velocity of approximately $35 \rm\,km\,s^{-1}$; the rotational axes of the stellar and gas components are also offset by approximately $30^\circ$ in projection. 

\eso{} also exhibits prominent interstellar Na\,D absorption, a tracer of neutral gas, which is redshifted by up to a few $\,\rm 100 \, km\,s^{-1}$, thereby tracing infalling gas, as explained in Section~\ref{sec: Na D absorption line analysis}.
Comparison of Figs.~\ref{fig: Halpha kinematics} and \ref{fig: Na D kinematics} suggests the narrow Na\,D component may be linked to the ionised gas disk traced by the narrow \ha{} emission, as both the LOS velocities and widths of the narrow-line gas and Na\,D are similar in the Western half of the galaxy.
The Na\,D absorption is also much deeper in the Northern regions of the galaxy, being entirely absent in the South (see Fig.~\ref{fig: Na D absorption profiles}). Similarly localised regions of redshifted interstellar Na\,D absorption have recently been observed in radio-active red geyser galaxies~\citep{Roy2021b}.

Together, our observations of the ionised and neutral phases of the ISM reveal an inflow of neutral gas streaming from the North, that is settling onto a disk that has become ionised in the interstellar radiation field, fuelling star formation and/or triggering jet activity.
Such inflow events are not unusual in ETGs; accretion is suspected to dominate the gas supply for $\approx 40\,\rm per\,cent$ of ETGs~\citep{Davis2011,Davis2019}.
Indeed, misalignments between the gas and stellar kinematics -- such as that observed in \eso{} -- are common in ETGs, and are believed to arise from gas accreting into the galaxy at an angle misaligned to the stellar axis of rotation~\citep{Bryant2019}.


\subsection{Merger or accretion event?}

We now investigate possible sources of the inflow into \eso{}.
It is highly unlikely that the gas is being accreted from a cosmic filament: as the group is only a projected distance of $\sim 1 \,\rm Mpc$ from the mega-structure to the North-East~\citepalias{Subrahmanyan2008}, any nearby filaments would preferentially accrete into this more massive structure instead of the small group housing \eso{}.
Likewise, the inflow is unlikely to be from a cooling flow in the IGM, as \eso{}, with an X-ray luminosity $L_{0.2-2\,\rm keV} \sim 10^{41}\,\rm erg\,s^{-1}$~\citepalias{Subrahmanyan2008}, lacks the bright X-ray emission associated with such cooling flows~\citep[e.g.,][]{McNamaraNulsen2012,Tremblay2018}.
We can also rule out stellar mass loss as the origin of the neutral gas, because the strength of the interstellar Na\,D absorption is non-uniform across the galaxy, being strongest in the North (Fig.~\ref{fig: Na D absorption profiles}). Moreover, its kinematics are highly distinct from that of the stars.
The most likely scenario is that the gas is being accreted from a neighbour, or is from a merger with a gas-rich, low-mass galaxy; indeed, the group containing \eso{} is unvirialised~\citepalias{Subrahmanyan2008}, facilitating interactions between group members.

\begin{figure}
	\centering
	\includegraphics[width=\linewidth]{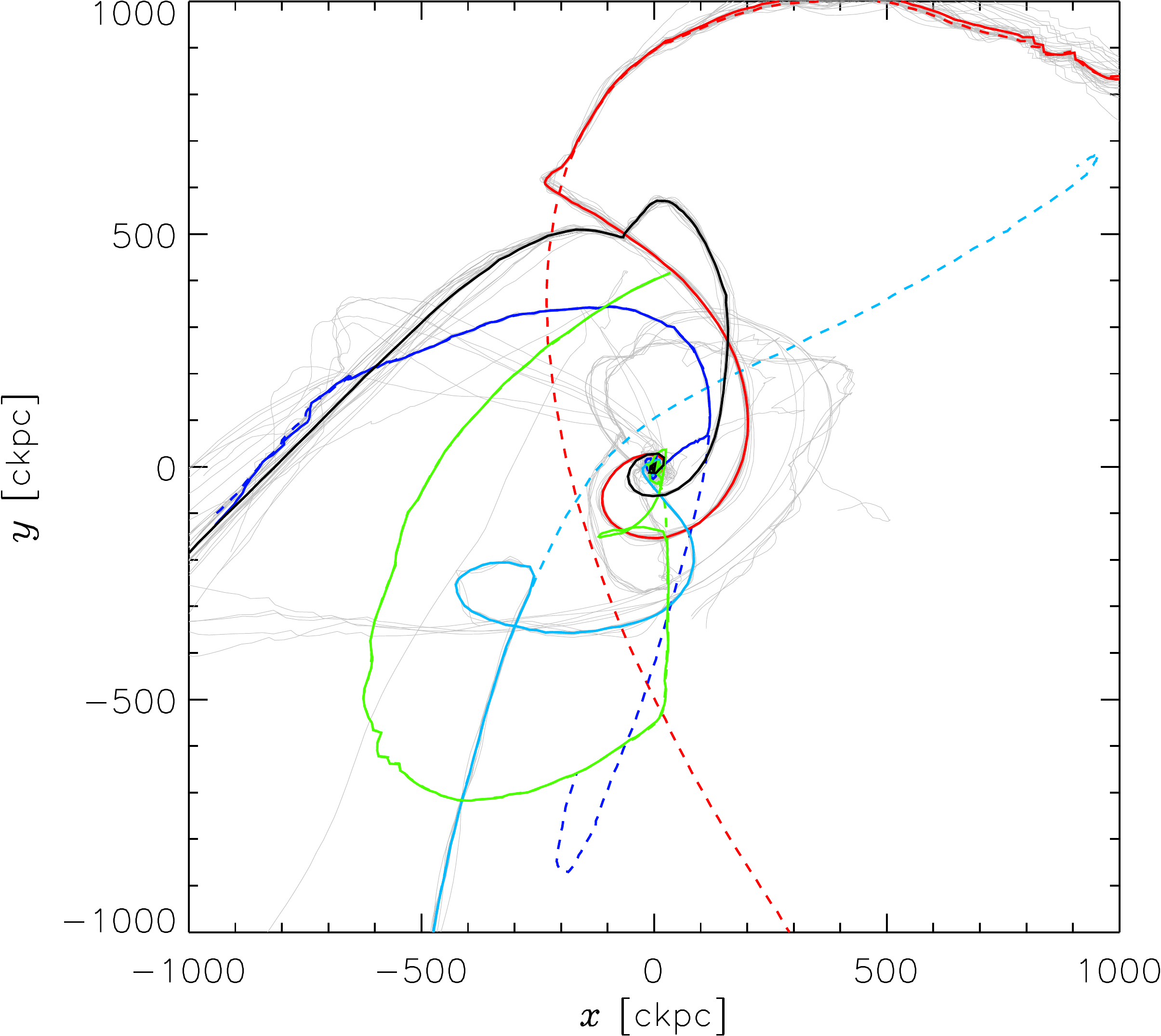}
	\caption{
		Trajectories of gas particles that are located in the simulated galaxy gaX0004, the closest analogue to \eso{}, at $z=0$, as well as those of nearby satellite galaxies.
		Gas particles and galaxies are followed back to $z=1$. Individual gas particles are shown in grey, important streams of gas are shown by solid coloured lines, and galaxy paths are represented by the dashed lines.
	}
	\label{fig:gas_tracks}
\end{figure}
\begin{figure}
	\centering
	\includegraphics[width=\linewidth]{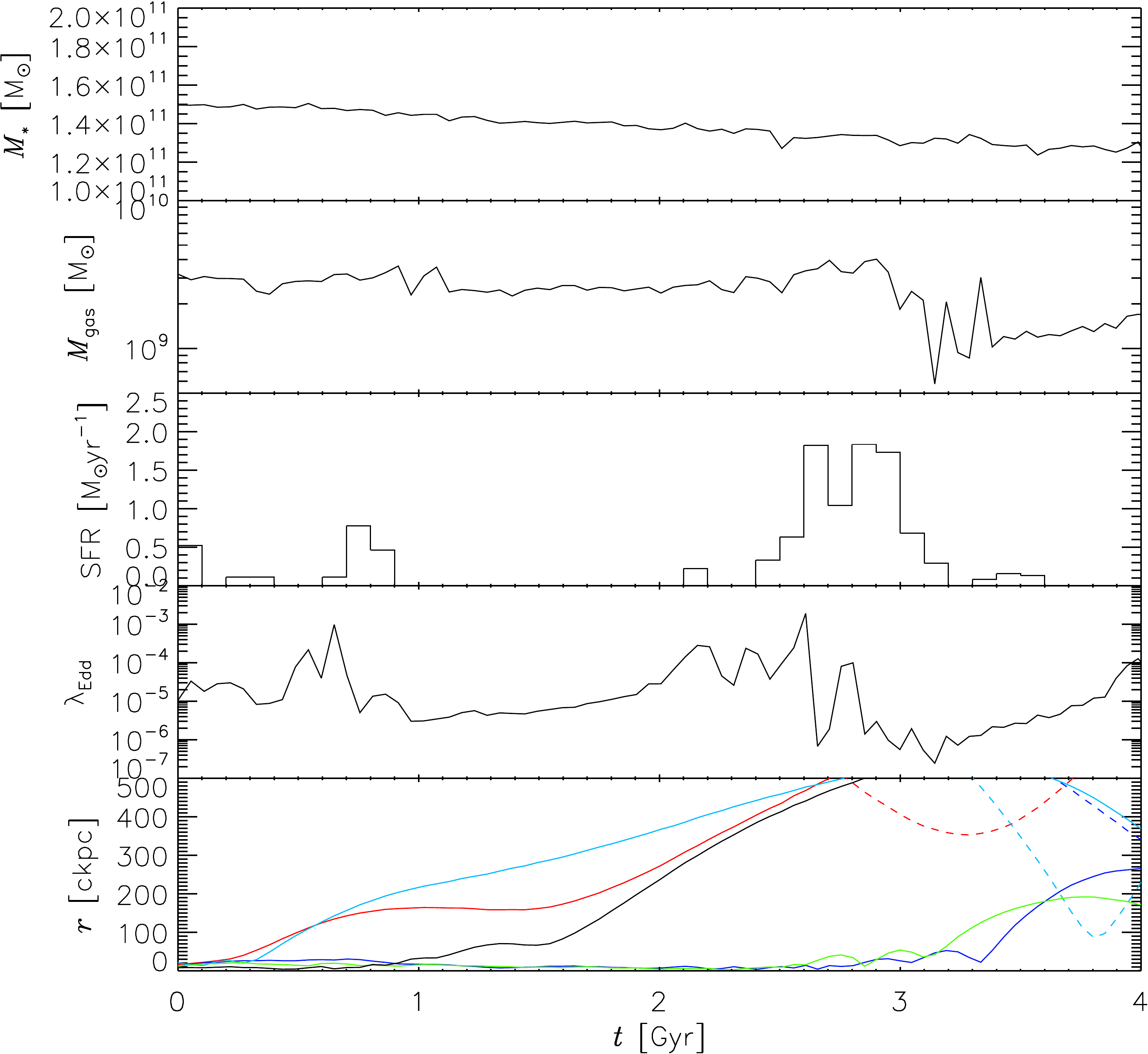}
	\caption{
		Properties of galaxy gaX0004 over 4 Gyr of lookback time since the present day.
		From top to bottom, panels show stellar mass, gas mass, star formation rate, Eddington ratio, and the distance of gas streams from gaX0004, where the colours in the bottom panel correspond to the tracks shown in the top panel.
	}
	\label{fig:ladder}
\end{figure}
To investigate these two scenarios, we analysed the formation history of a massive galaxy similar to \eso{} in the cosmological simulation of Kobayashi \& Taylor et al., \textit{in prep}.
The simulation is similar to that introduced in \citet{Taylor&Kobayashi2015a}, using the same initial conditions comprising $240^3$ particles of each of gas and dark matter in a cubic box $25\,h^{-1}$ Mpc (comoving) per side.
The AGN feedback scheme includes formation from dense, metal-free gas, gas accretion assuming the Bondi-Hoyle formalism, isotropic thermal feedback, and black hole mergers~\citep{Taylor&Kobayashi2014}.
The simulation adopts the updated chemical evolution and stellar feedback model of \citet{Kobayashi2020b}, which tracks all individual elements from H to Ge, and includes prescriptions for chemical enrichment and feedback from core-collapse supernovae and AGB winds \citep{Kobayashi2020b}, supernovae Ia \citep{Kobayashi&Nomoto2009,Kobayashi2020a}, as well as a metallicity-dependent hypernova fraction \citep{Kobayashi2020b}.
The simulation assumes the following cosmological parameters: $h=0.68$; $\Omega_{\rm M} = 0.31$; $\Omega_\Lambda = 0.69$; $\Omega_{\rm b}=0.048$.

We identified galaxies at $z=0$ in this simulation with properties similar to \eso{}.
The closest match, which we will refer to as gaX0004, has $M_* = 1.5\times10^{11}\,\rm M_{\odot}$, $M_{\rm gas} = 3.2\times10^9\,\rm M_{\odot}$, $M_{\rm BH} = 2.4\times10^8\,\rm M_{\odot}$, $\rm SFR = 0.52\,\rm M_{\odot}\,\rm yr^{-1}$, and $\dot{M}_{\rm BH} = 5.7\times10^{-5}\,\rm M_{\odot}\,{\rm yr}^{-1}$.
By matching particle IDs in different snapshots, we follow the gas particles that exist in gaX0004 at $z=0$ back through the simulation to understand their origin.

Fig.~\ref{fig:gas_tracks} shows the trajectories of these gas particles from $z=1$ to the present, corresponding to a span of approximately 8 Gyr.
Some of these gas particles are previously associated with low-mass satellites of gaX0004; the motions of the centres of mass of these particles is shown by the solid coloured lines, and the trajectories of their former host galaxies are shown by the equivalently coloured dashed lines.
In all four cases the gas is stripped from its galaxy before being accreted by gaX0004, and only one of the satellite galaxies (green in Fig.~\ref{fig:gas_tracks}) has merged with gaX0004 by $z=0$\footnote{Although not important in this context, we note that the gas is removed from these satellites due to interactions with an AGN-driven outflow from gaX0004 itself. Further details can be found in Taylor et al. (\textit{in prep.}).}.
Also shown in black is the path of a coherent, bound mass of gas that falls onto gaX0004, but which is not associated with any galaxy.

Fig.~\ref{fig:ladder} shows the stellar mass, gas mass, SFR and Eddington ratio $\lambda_{\rm Edd}$ of gaX0004, as well as the galactocentric distance of the gas streams shown in Fig. \ref{fig:gas_tracks}.
As expected for a massive galaxy at low redshift, the stellar mass increases only slightly over this period, and both the gas fraction and SFR remain low.
However, each gas accretion or merger event is accompanied by first a moderate increase in the SFR, followed by an increase in $\lambda_{\rm Edd}$, due to the additional time taken for gas to fall towards the central black hole. In particular, the most recent accretion event has triggered SFRs of approximately $0.5\,\rm M_\odot \,yr^{-1}$ and Eddington ratios of $\lambda_{\rm Edd} \sim 10^{-5}$.

These simulations therefore show that both merger events and accretion from either nearby galaxies or the IGM are capable of triggering both star formation and AGN activity at the levels observed in \eso{}.
Deep \hi{} observations of the group that could reveal any neutral streams associated with neighbouring galaxies may distinguish whether the gas originated in a satellite galaxy or the IGM.
It may also be possible to determine the origin of the gas by measuring its chemistry: whilst the stripped gas has approximately solar metallicity, the material from the IGM is enriched to less than one tenth the solar value.
Elemental abundance ratios tell a similar story: the IGM gas is significantly $\alpha$ enhanced, with ${\rm [O/Fe]} = 1.91$ and ${\rm [Mg/Fe]} = 2.35$, compared to $\sim 0.5$ and $\sim 1$, respectively, for the stripped gas.
Unfortunately we cannot estimate the metallicity of the inflow without first constraining the total gas mass; similarly we lack the S/N to accurately estimate the metallicity of either kinematic component in the ionised gas.

Interestingly, as shown in Fig.~\ref{fig: stellar age map}, the most recent star formation event began roughly 50--60\,Myr ago, whilst the estimated age of the jet is a few Myr (Riseley et al., \textit{in prep.}), implying a time delay of a few tens of Myr between triggering of star formation and the AGN, which is broadly consistent with the delays visible in Fig.~\ref{fig:ladder}. However, we note that our stellar age estimates are unlikely to be accurate to within a few 10s of Myr, and so we cannot confidently estimate the length of this delay from our observations.


\subsection{What is the evolutionary significance of the inflow?}

\eso{} is undergoing an epoch of restarted jet activity as evidenced by pc-scale jet structures in the nucleus, as well as multiple breaks in the radio-frequency SED, with an estimated age of a few Myr (Riseley et al., \textit{in prep.}). 

The comparable timescales of the recent star formation and jet activity may indicate that the inflowing material is responsible for the current epoch of AGN activity.
Inflows of neutral gas traced by \hi{} have been reported in radio galaxies~\citep{vanGorkom1989} and may fuel AGN activity; \citet{Sato2009} speculate that these same inflows fuelling ``maintenance-mode feedback'' may also be observed in Na\,D. 
Indeed, localised regions of redshifted interstellar Na\,D absorption have recently been detected in radio-active red geyser galaxies alongside outflows and low star formation rates, suggesting a link between the inflow, jet activity and feedback~\citep{Cheung2016,Roy2021a,Roy2021b}.

Although the estimated inflow rate in \eso{} ($\sim 1-3 \,\rm M_\odot \, yr^{-1}$) is much higher than those detected in \hi{}~\citep[$\sim 0.01 \,\rm M_\odot \, yr^{-1}$;][]{vanGorkom1989}, it is within the range of measured Na\,D-detected inflow rates in radio-active red geyser galaxies~\citep[$\sim 0.02 - 5 \,\rm M_\odot \, yr^{-1}$;][]{Roy2021b}. 
Meanwhile, \citet{Krug2010} detected Na\,D inflows of up to approximately $0.7 \rm \, M_\odot \, yr^{-1}$ in roughly a third of their sample of infrared-faint Seyfert galaxies.

It is therefore unclear whether the inflowing gas will form an accretion disk, thereby triggering more powerful Seyfert-like AGN activity via radiatively efficient accretion, or whether it will trigger radiatively-inefficient accretion conducive to powerful jet activity.
Radio galaxies have been observed to maintain their LERG status even after the arrival of cold gas~\citep{Janssen2012}. Indeed, the observed \forb{O}{iii} luminosity implies radiatively inefficient accretion, with an estimated accretion rate of $\sim 10^{-2}\,\rm M_\odot$ (Section~\ref{subsec: Eddington ratio}), although the pc-scale jets may not have yet grown to a size where they can interact with the kpc-scale gas probed by our WiFeS observations. The future state of the radio source in \eso{} is therefore uncertain.

\citet{Davis2011} predict that newly accreted gas that is misaligned with the existing gas disk will rapidly lose its angular momentum and fall towards the centre of the galaxy due to dissipative collisions between the two components. 
The inflow in \eso{} is misaligned with the existing gas disk, as indicated by the presence of two misaligned kinematic components in the ionised gas.
Our observations are therefore consistent with a scenario in which the gas is unstable, potentially fuelling star formation and/or nuclear activity as it falls into the centre of the galaxy.
The apparent paucity of multiple, misaligned gas disks in ETGs may reflect the short-lived nature of such phenomena.

\section{Summary \& conclusions}

We presented our study of \eso{}, a recently restarted giant radio galaxy, using optical integral field spectroscopy from the WiFeS spectrograph. Our findings are summarised as follows:
\begin{enumerate}
	\item Unusually for a slow rotator ETG, \eso{} harbours a young ($\lesssim 10\,\rm Myr$) stellar population concentrated in the North-West regions of the galaxy with an estimated mass of $10^{7.6}\,\rm M_\odot$ and an ongoing SFR of $0.4 \,\rm M_\odot \, yr^{-1}$. 
	\item \eso{} also exhibits highly unusual ionised gas kinematics: the \ha{} emission traces two kinematically distinct gas disks, characterised by narrow ($\sigma_{\rm H\upalpha} \lesssim 100\,\rm km\,s^{-1},\, v_{\rm rot} \approx 300 \,\rm km\,s^{-1}$) and broad ($\sigma_{\rm H\upalpha} \gtrsim 150\,\rm km\,s^{-1},\,v_{\rm rot} \approx 100 \,\rm km\,s^{-1}$) emission lines respectively.
	\item Prominent redshifted interstellar Na\,D absorption is present in the North, tracing an inflow of neutral gas with an estimated inflow rate of $1 - 3 \,\rm M_\odot \, yr^{-1}$. Both broad ($\sigma_{\rm Na\,D} \gtrsim 200\,\rm km\,s^{-1}$) and narrow ($\sigma_{\rm Na\,D} \lesssim 100\,\rm km\,s^{-1}$) components were detected; the narrow component is purely interstellar in origin, whereas the broad component may be contaminated by stellar absorption. Whilst the narrow component appears to be linked to the ionised gas disk traced by the narrow \ha{} emission, the nature of the broad component is less clear.
\end{enumerate}

We propose that \eso{} is undergoing a minor merger or gas accretion event. The corresponding inflow of gas is occurring from the North, at an angle misaligned with the axis of stellar rotation. The neutral gas has settled into a disk, and may have triggered the observed star formation, resulting in the narrow-line ionised gas disk. This configuration is unstable, as evidenced by kinematic offset between the stars and gas, and will eventually dissipate, resulting in an inflow of gas towards the centre of the galaxy. The presence of pc-scale jets indicates that at least some of this gas has already made its way to the nucleus, triggering the next epoch of jet activity. This fuelling-starburst-AGN sequence is consistent with the evolution of giant galaxies observed in cosmological simulations.



%% file: appendix.tex
\section{Example Na\,D absorption profile fits}\label{sec: appendix: example NaD fits}

\begin{figure*}
	\centering
	\includegraphics[width=1\linewidth]{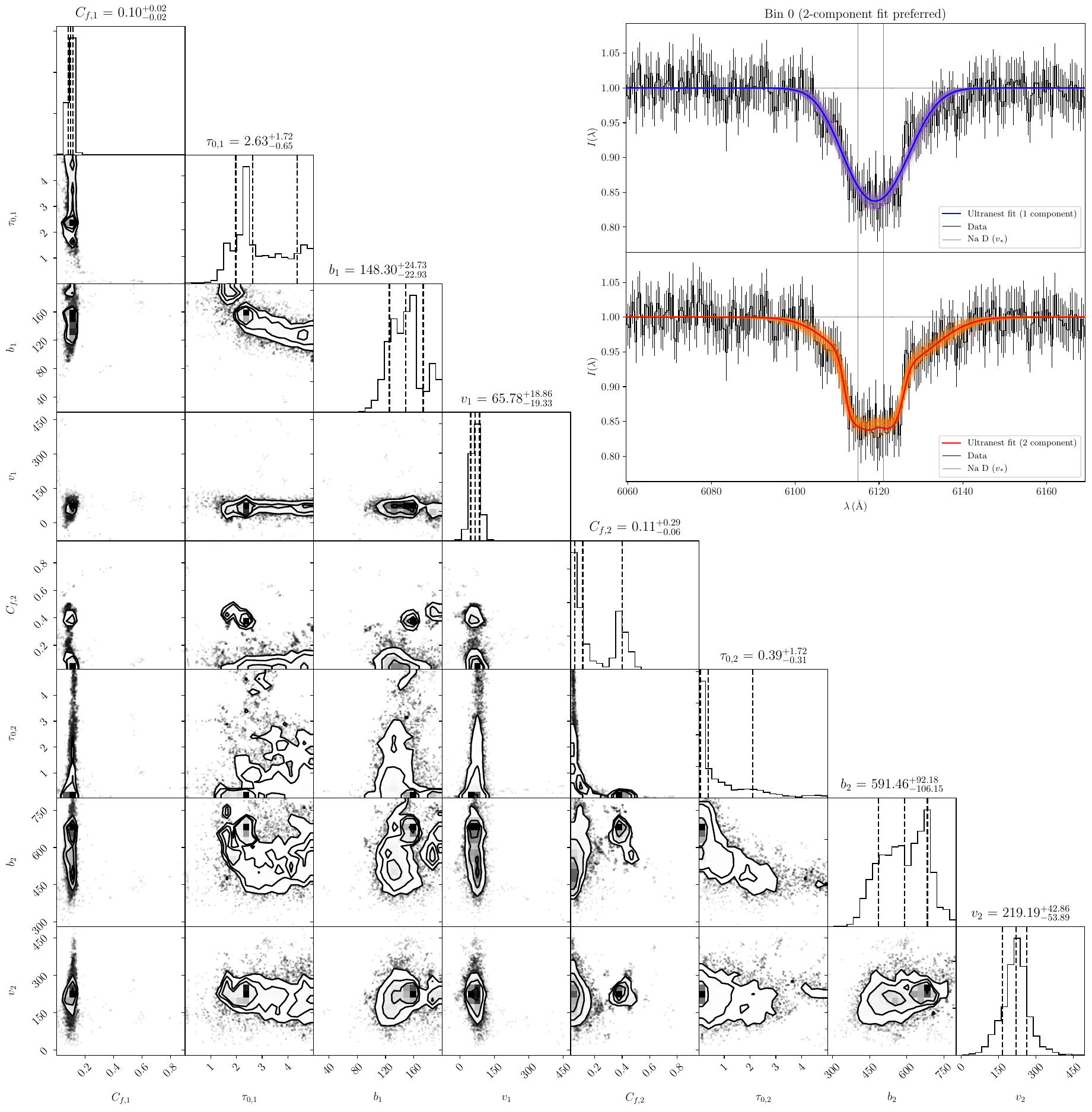}
	\caption{Corner plot showing the posterior PDFs sampled using \textsc{Ultranest} corresponding to the Na\,D absorption profile fit from bin 0 (inset). In the corner plot, the vertical dashed lines indicate 16th, 50th and 84th percentile confidence intervals in the marginalised posterior PDFs. In the inset plot, the upper and lower panels show the fit assuming one and two kinematic components respectively. The pale purple and orange lines show models generated from randomly selected subsets of parameters sampled by \textsc{Ultranest}.}
	\label{fig: appendix: example NaD fit (bin 0)}
\end{figure*}

\begin{figure}
	\centering
	\includegraphics[width=1\linewidth]{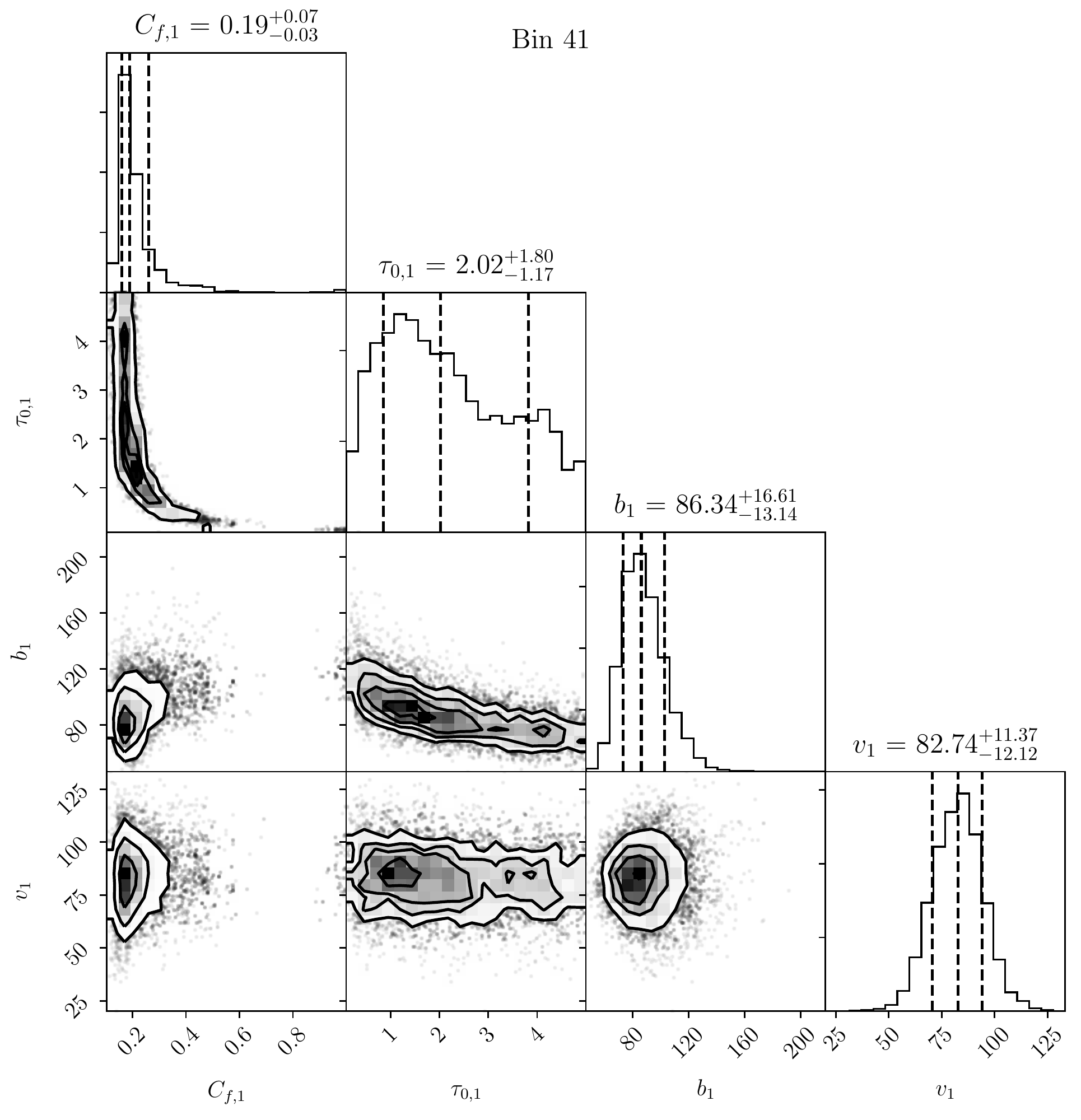}
	\includegraphics[width=1\linewidth]{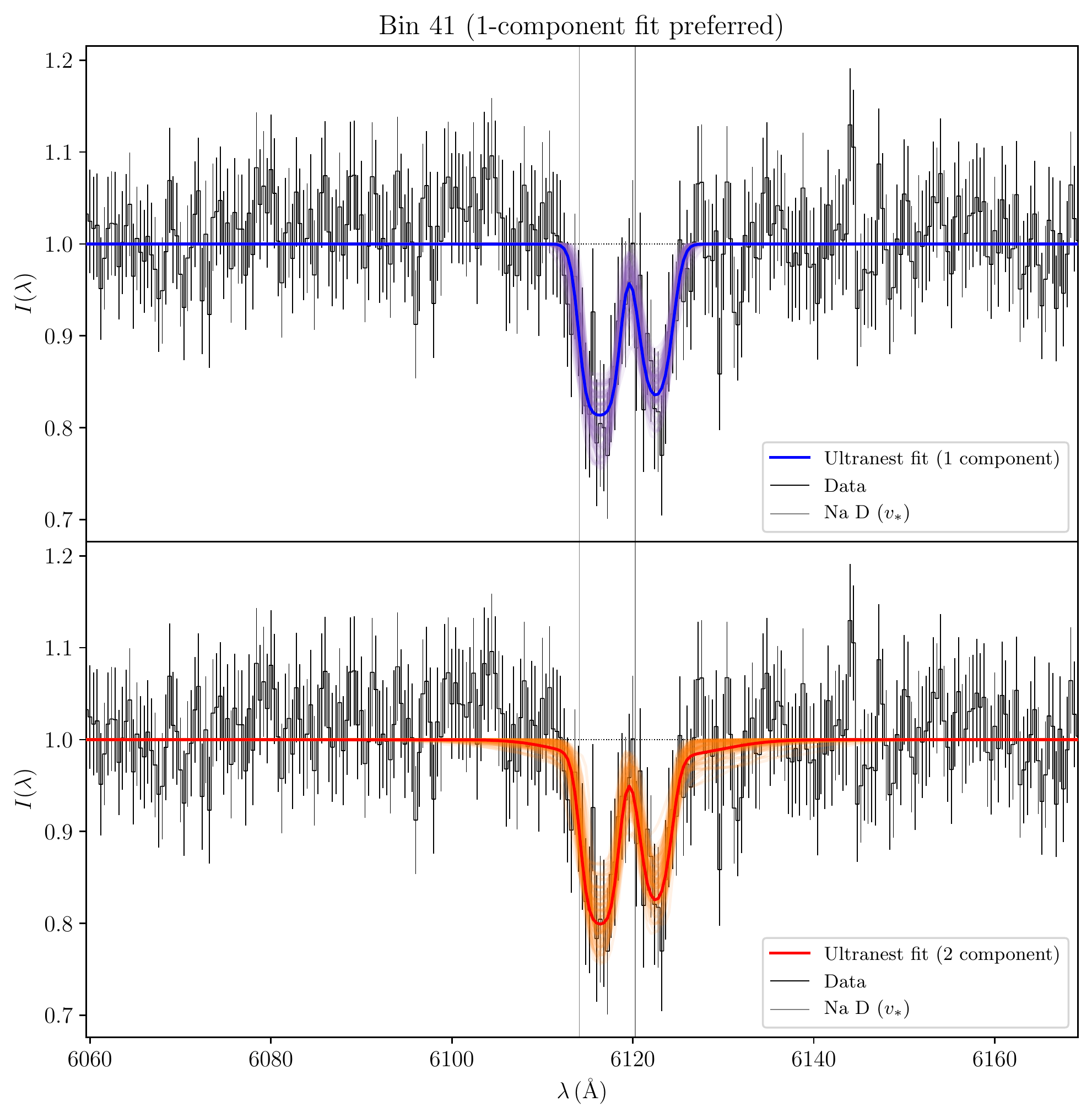}
	\caption{Same as Fig.~\ref{fig: appendix: example NaD fit (bin 0)} for bin 41.}
	\label{fig: appendix: example NaD fit (bin 41)}
\end{figure}

Na\,D profile fits and corner plots for two bins, generated using the \textsc{corner} package for \textsc{python}~\citep{Foreman-Mackey2016}, showing the posterior parameter distributions sampled by \textsc{Ultranest}, are shown in Figs.~\ref{fig: appendix: example NaD fit (bin 0)} and \ref{fig: appendix: example NaD fit (bin 41)}. In the former, a two-component fit is preferred, whilst in the latter only a single component is present. 

\begin{figure}
	\includegraphics[height=0.7\linewidth]{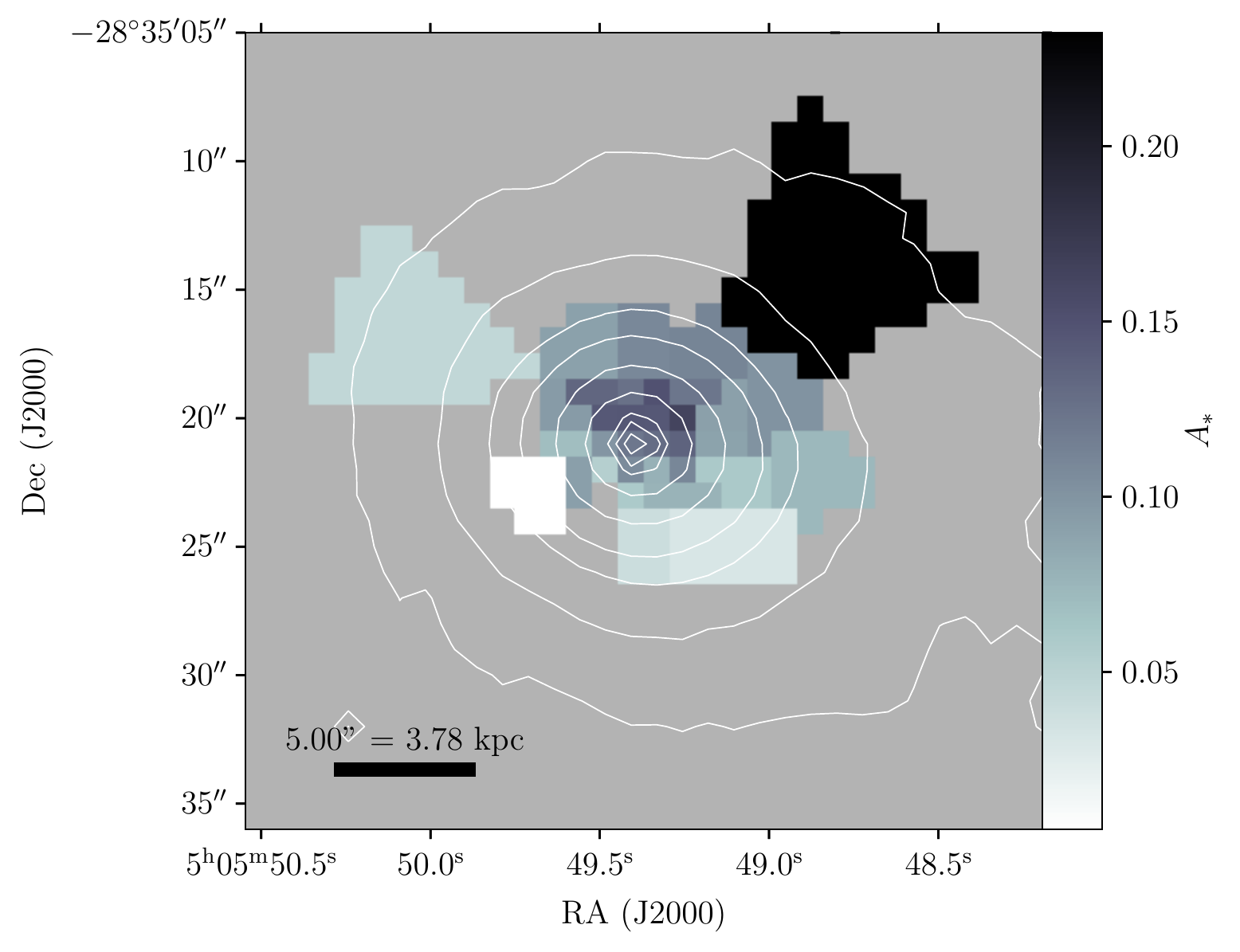}
	\caption{Amplitude $A_*$ of the best-fit residual stellar absorption component in each bin, assuming the broad component is stellar in origin. The contours are as in Fig.~\ref{fig: stellar age map}.}
	\label{fig: Na D stellar amplitude}
\end{figure}

\section{Accounting for a possible stellar component in the Na\,D absorption}\label{sec: appendix: residual stellar Na D absorption}

To investigate whether the broad residual Na\,D absorption is stellar in origin, we repeated our analysis detailed in Section~\ref{sec: Na D absorption line analysis} using an additional Gaussian absorption component to account for residual stellar absorption.
In each bin, a single Gaussian profile was first fitted to the Na\,D absorption feature in the best-fit stellar continuum obtained from our \ppxf{} analysis using \textsc{mpfit}.
We then re-ran our absorption profile fit in each bin, substituting a single Gaussian for the broad interstellar component as follows. 
The absorption profile in each bin was inspected to determine the ideal combination of fitted components: either a single narrow interstellar Na\,D component, a single stellar Gaussian component, or both, or no profile at all.
In those bins where a stellar component was included, a Gaussian absorption component was fitted such that the width and central wavelength of the Gaussian was fixed to that of the stellar template, and only the amplitude $A_*$ was allowed to vary. 

The amplitude of the residual stellar absorption component $A_*$ in each bin is shown in Fig.~\ref{fig: Na D stellar amplitude}. 
The substitution of the stellar component slightly increased the reduced-$\chi^2$, but otherwise produced a satisfactory fit to the data in most bins. However, $A_*$ increases significantly from South to North, as shown in Fig.~\ref{fig: Na D stellar amplitude}.
Were the broad component purely stellar in origin, $A_*$ would be relatively uniform across the galaxy; it is therefore unlikely that the broad component is purely stellar in origin.